\documentclass{article}
\usepackage{arxiv}

\usepackage{amsmath,amsthm,amssymb,amscd}
\usepackage[all,cmtip]{xy}
\usepackage{diagxy}
\usepackage[utf8]{inputenc} % allow utf-8 input
\usepackage[T1]{fontenc}    % use 8-bit T1 fonts
\usepackage{hyperref}       % hyperlinks
\usepackage{url}            % simple URL typesetting
\usepackage{booktabs}       
\usepackage[linesnumbered,boxed]{algorithm2e}
\usepackage{amsfonts}       % blackboard math symbols
\usepackage{nicefrac}       % compact symbols for 1/2, etc.
\usepackage{microtype}      % microtypography
\usepackage{lipsum}
\usepackage{mathrsfs}
\usepackage{fancyhdr}
\usepackage{color}
\usepackage{subfig}
\usepackage{graphicx}

\newtheorem*{definition}{Definition}
\newtheorem{lemma}{Lemma}
\newtheorem{proposition}{Proposition}
\newtheorem{corollary}{Corollary}
\newtheorem{theorem}{Theorem}
\newtheorem*{remark}{Remark}

\newcommand{\id}{{\rm{id}}}

\newcommand{\bbZ}{\mathbb{Z}}
\newcommand{\bbR}{\mathbb{R}}
\newcommand{\bbN}{\mathbb{N}}

\newcommand{\bfa}{{\mathbf a}}
\newcommand{\bfb}{{\mathbf b}}
\newcommand{\bfe}{{\mathbf e}}
\newcommand{\bfu}{{\mathbf u}}

\newcommand{\bfx}{{\mathbf x}}

\newcommand{\bfp}{{\mathbf p}}

\graphicspath{{figures/}}

\title{A Multi-parameter Persistence Framework for Mathematical Morphology}

\author{
  Yu-Min Chung \\ 
  Department of Mathematics and Statistics\\
  University of North Carolina at Greensboro\\
  Greensboro, USA  \\
  \texttt{y\_chung2@uncg.edu} \\
  \AND
  Sarah Day \\
  Department of Mathematics\\
  William \& Mary, Williamsburg \\
  Virginia 23185, USA \\
  \texttt{sldayx@wm.edu} \\
  \AND
  Chuan-Shen Hu \\ 
  Department of Mathematics\\
  National Taiwan Normal University\\
  Taipei, Taiwan \\
  \texttt{peterbill26@hotmail.com}
}

\begin{document}
\maketitle

\begin{abstract}
The field of mathematical morphology offers well-studied techniques for image processing.  In this work, we view morphological operations through the lens of {\em persistent homology}, a tool at the heart of the field of topological data analysis.  We demonstrate that morphological operations naturally form a multiparameter filtration and that persistent homology can then be used to extract information about both topology and geometry in the images as well as to automate methods for optimizing the study and rendering of structure in images.  For illustration, we apply this framework to analyze noisy binary, grayscale, and color images.     
\keywords{Mathematical morphology \and Topological data analysis \and Multi-parameter filtration \and Persistent homology \and Salt and pepper noise}
\end{abstract}

% keywords can be removed

\section{Introduction}
\label{sec:intro}
Computational topology and the field of topological data analysis offer powerful tools for analyzing structure in data ~\cite{MischaikowNanda,zomorodian2005computing,ghrist2008,edelsbrunner2008persistent,gunther2012efficient}. Persistent homology, in particular, has offered a means of measuring topological features across a {\em filtration}, or sequence of structures built from the data.  A one-filtration is a collection of nested sets where the inclusion map enables the tracking of topological information from one set to the next. A {\em multifiltration} extends this notion to an indexed collection of sets satisfying an inclusion relation between a pair of sets whenever their indices are related under a specified partial order.  This allows for the construction of many structures related to image or point cloud data, where appropriate inclusion relationships allow for the tracking of topological information across the structures.

To date, most studies using {\em multiparameter persistence}, the extension of persistent homology to multifiltrations, have focused on point cloud data rather than the cubical/digital image data we study here \cite{carlsson2010computing,carlsson2009theory,lesnick2019computing,corbet2019kernel,vipond2020multiparameter,carriere2020multiparameter}.  {A standard filtration for grayscale digital images is the sublevel set filtration obtained via thresholding.  In this work, we use morphological operations to construct new filtrations, yielding erosion~\eqref{equ:erosion filtration}, dilation~\eqref{equ:dilation filtration}, opening~\eqref{equ:opening filtration}, and closing filtrations~\eqref{equ:closing filtration} and their variants ~\eqref{equ:wth filtration}, \eqref{equ:bth filtration}, \eqref{equ:sth filtration}. We then show that under mild assumptions, combinations of these operations form multiparameter filtrations.}  This establishes a multifiltration framework for the analysis of digital images where features appear at different spatial scales, including noisy images in which the noise is smaller in spatial scale than the underlying structure we wish to uncover.  We then demonstrate that it is possible to use this framework and persistent homology to extract information about underlying structure in the images as well as to automate the production of a denoised image.  

Digital images may naturally be thought of as functions on $\bbZ^m$, with sets of pixels for regions of interest in the image given as subsets of $\bbZ^m$.   In ~\cite{10.1117/1.1789986}, the authors provide an interesting method for smoothing shapes of objects in $\bbZ^m$, with $m = 2,3$, which preserves homotopy structure. To achieve this, the authors give the definition of homotopic equivalence of discrete images and construct homotopic thinning/thickening operations for shape smoothing. However, in image denoising tasks, one often aims to remove small scale features that arise due to noise in the image, thus sometimes dramatically changing the topology of the image.  Therefore, in our approach, we do not impose homotopic equivalence and instead adopt a goal of intentionally changing the homological type of rendered structures in order to optimize topological and geometric accuracy by removing features most likely due to noise.

Filtrations and persistence lend themselves well to automation.  In \cite{ChungDay}, Chung and Day used persistent homology to track structure in the sublevel set filtrations, developing an automated method for extracting topological measurements and thresholding grayscale images.  In this work, we focus on building an algebraic topological framework for the application of the  morphological image processing operations of erosion, dilation, opening, and closing.  These are well-developed operations for cleaning images by removing small scale features while keeping the remainder of the image relatively constant \cite{Soille2003,Najman-Mathematical-Morphology,serra1984image}.  When combining morphological operations, the dimension of the constructed multiparameter filtration grows rapidly in the numbers of operations and utilized structuring elements (see Section~\ref{sec:background}).  Furthermore, thresholding may be combined with opening and closing to form a yet larger multifiltration for studying grayscale and, by extension, color images.  As illustration, {in Section~\ref{sec:denoising algorithm}, we use opening, closing, and thresholding to construct a multifiltration that we use for denoising images with salt and pepper noise, providing sample results for binary, grayscale, and color images.}

In what follows, we introduce necessary definitions and properties for morphological operations (Section~\ref{sec:background}) and persistent homology (Section~\ref{sec:ph}).  We then use morphological operations and thresholding to build multifiltrations in Section~\ref{sec:multiparameter filtrations}, presenting our main results in Theorem~\ref{thm:multifiltration-peter} and Corollary~\ref{thm:multifiltration2-peter}.  This presents the necessary framework to combine morphological operations and, if appropriate, thresholding, in a single multiparameter filtration for analyzing digital data.   We show some illustrative uses of these filtrations in Section~\ref{sec:denoising algorithm}.

\section{Background on Mathematical Morphology}
\label{sec:background}
Mathematical morphology is a field that provides theoretical and practical techniques for processing digital images ~\cite{4767941,SILVA202033,sonka2014image,heijmans1998mathematical,cousty:tel-01862637}. In later sections, we will focus on using morphological operations to measure and track topological features in images.  Here, we focus on establishing properties of the operations that are necessary to this topological approach.  Two morphological operations that will be of particular interest in what follows are {\em dilation} and {\em erosion}. In a binary image, {\em dilation} enlarges features in the pixel subset for a specified value (e.g. $1$, $255$, or `black') while {\em erosion} erases small, isolated features in the pixel subset (see \cite{Soille2003,Najman-Mathematical-Morphology,serra1984image} and references therein).

We denote integers, natural numbers, and real numbers by the standard notation $\mathbb{Z}$, $\mathbb{N}$, and $\mathbb{R}$, respectively. We use $\mathbb{Z}_{\geq 0}$ to denote $\mathbb{N} \cup \{ 0\}$, the set of all non-negative integers. The symbol $\mathbb{R}_{\geq 0}$ represents the set of all non-negative real numbers. Elements in $\mathbb{Z}^m$ are denoted by boldface letters \textit{e.g.} $\mathbf{u} \in \mathbb{Z}^m$ to distinguish vectors and scalars.  We use this notation to build towards formalizing operations on digital images, which we define as follows.

\begin{definition}
[Section 1.1.2.1.~\cite{Najman-Mathematical-Morphology}, p. 6]
\label{def:image}
Let $m \in \mathbb{N}$ be a positive number, an $m$\textit{-dimensional (digital) image} on \textit{pixel/voxel set} $P\subseteq \mathbb{Z}^m$ is a non-negative function $g : P \rightarrow \mathbb{R}_{\geq 0}$. If the range of the function $g$ is $\{0, 1\}$, then $g$ is called a \textit{binary image}. Otherwise, we refer to $g$ as a \textit{grayscale image}. The set of all images on $P$, denoted $\mathcal{I}_P$, is defined as $\mathcal{I}_P =\{g : P \rightarrow \mathbb{R}_{\geq 0}\}.$  
\end{definition}

In practice, one usually considers a rectangular image whose domain can be expressed as
\begin{equation*}
    P = \bbZ^m \cap \left( \prod_{i = 1}^m [a_i, b_i] \right)
\end{equation*}
where $a_i \leq b_i$. 

The following discussion focuses primarily on binary images $g : P \rightarrow  \{ 0, 1\}$, where a value of $1$ means the pixel is white and a value of $0$ means the pixel is black,  
and $8$-bit grayscale images $g:P \rightarrow \{ 0,1, ..., 255\}$ where $225$ is white and $0$ is black. When feasible, we consider general images $g : P \rightarrow \mathbb{R}_{\geq 0}$ so that properties and theorems stated in the paper hold in this general case.   This setting is also convenient when considering re-scaling of pixel values of images with different range sets. 

We now establish the following partial order on the space of images and define image and preimage subsets.  Given two images $f, g : P \rightarrow \mathbb{R}$, we say that 
$$f \leq g \text{ if and only if } f(\bfp) \leq g(\bfp)~ \text{ for all } \bfp \in P.$$ 
For functions $f : S \rightarrow T$, $A \subseteq S$ and $B \subseteq T$, the sets $f(A) := \{ f(a) \ | \ a \in A \}$ and $f^{-1}(B) := \{ s\in S \ | \ f(s) \in B \}$ are the {\em image of $A$} % \subseteq S$ 
and the {\em preimage of $B$} under $f$.  % \subseteq T$ via $f$. 
As an abbreviation, if $B = \{ b \}$ is a singleton set, we write $f^{-1}(b)$ instead of $f^{-1}(\{b \})$ to denote $f^{-1}(B)$.

\begin{definition}
[Section 3.1~\cite{Soille2003} p. 64]
For $m \in \mathbb{N}$, a \textit{structuring element} is a specified finite set $B$ satisfying $\mathbf{0} \in B \subseteq \mathbb{Z}^m$. A structuring element $B$ is \textit{symmetric} if $B = -B := \{ -\bfx \ | \ \bfx \in B \}$.
\end{definition}

\begin{remark}
In mathematical morphology, structuring elements defined here are called \textit{flat} structuring elements ~\cite{Soille2003}. In certain applications, a \textit{non-flat} structuring element $\mathcal{B}$ is defined as a function from a finite subset $B$ of $\bbZ^m$ to $\bbR_{\geq 0}$ which records weighted values for elements in $B$. In this case, a flat structuring element can be viewed as a characteristic function $\chi_B$ on a finite set $B \subseteq \bbZ^m$. In this paper, all structuring elements we consider are flat.  
\end{remark}

Structuring elements will be used to define local windows over which pixel values are considered during processing operations. This requires the {\em Minkowski sum} and {\em difference} for subsets $A,~B \subseteq \mathbb{Z}^m$, defined respectively as  
\begin{equation*}
    \begin{split}
        A + B &:= \{ \bfa + \bfb \ | \ \bfa \in A, \bfb \in B \}, \\
        A - B &:= \{ \bfa - \bfb \ | \ \bfa \in A, \bfb \in B \}.
    \end{split}
\end{equation*}
If either $A = \{ \bfa \}$ or $B = \{ \bfb \}$ are sets of singleton points, we would simply use $\bfa + B$, $\bfa - B$, $A + \bfb$ or $A - \bfb$ rather than $\{ \bfa \} + B$, $\{ \bfa \} - B$, $A + \{ \bfb \}$ or $A - \{ \bfb \}$ to denote the Minkowski sum or difference of $A$ and $B$s.

Since ${\bf0} \in B$, one may think of $\bfx +B$ as the {\em $B$-neighborhood} of $\bfx$ in $\mathbb{Z}^m$. If $g \in \mathcal{I}_P$ is a binary image and $g(\bfx + B) = \{ 1 \}$, then the $B$-neighborhood of $\bfx$ is contained in the white region of the image. On the other hand, if $g(\bfx + B) = \{ 0, 1 \}$, then the $B$-neighborhood of $\bfx$ intersects both the white and the black sets in the image.

In this work, we consider four fundamental morphological operations: {\em erosion}, {\em dilation}, {\em opening}, and {\em closing}.  We next review their formal definitions.
\begin{definition}
[Equations (1.6) and (1.7) in~\cite{Najman-Mathematical-Morphology} p. 10]
For $g \in \mathcal{I}_P$ and structuring element $B \subseteq \mathbb{Z}^m$, the \textit{erosion} of $g$ via $B$ is an image $\epsilon_B(g) \in \mathcal{I}_P$ defined by
\begin{equation}
\label{Eqaution : Erosion}
    \epsilon_B(g)(\bfx) = \min g\bigg((\bfx + B) \cap P\bigg) = \min \{ g(\bfx+\bfb) \ | \ \bfb \in B, \bfx+\bfb \in P \}.
\end{equation}
%{\color{red}Why $\bfx$ and $x$?}
Similarly, the \textit{dilation} of $g$ via $B$ is an image $\delta_B(g) \in \mathcal{I}_P$ defined by
\begin{equation}
\label{Eqaution : Dilation}
    \delta_B(g)(\bfx) = \max g\bigg((\bfx - B) \cap P\bigg) = \max \{ g(\bfx-\bfb) \ | \ \bfb \in B, \bfx-\bfb \in P \}.
\end{equation}
Since $\mathbf{0} \in B$ and $B$ is finite, $(\mathbf{x} + B) \cap P$ and $(\mathbf{x} - B) \cap P$ are non-empty, finite sets whenever $\bfx \in P$. Therefore, $\epsilon_B(g)$ and $\delta_B(g)$ are well-defined.
\end{definition}

\begin{remark}
Observe that if $B \subseteq \bbZ^m$ is symmetric, then \eqref{Eqaution : Dilation} is equivalent to
\begin{equation}
\label{Eqaution : Dilation for symmetric B}
    \delta_B(g)(\bfx) = \max g\bigg((\bfx + B) \cap P\bigg) = \max \{ g(\bfx+\bfb) \ | \ \bfb \in B, \bfx+\bfb \in P \}.
\end{equation}
\end{remark}

Erosion and dilation may now be composed to define the operations of {\em opening} and {\em closing}.
\begin{definition}
[Section 1.2.1~\cite{Najman-Mathematical-Morphology}, p. 12]
Let $P,B\subseteq \mathbb{Z}^m$ be a pixel set and  structuring element respectively. Then \textit{opening} and \textit{closing} operations via $B$, denoted by $O_B$ and $C_B$ respectively, are functions $O_B, C_B: \mathcal{I}_P \rightarrow \mathcal{I}_P$ defined as
\begin{equation}
\label{def:open and close}
    O_B = \delta_B \circ \epsilon_B \ \ {and} \ \ C_B = \epsilon_B \circ \delta_B.
\end{equation}
\end{definition} 
The opening and closing operators may be used to remove structure that is smaller than the scale prescribed by $B$ while minimizing distortion of larger scale features \cite{dougherty1989image,fletcher2005texture,marcosa2013probabilistic,Soille2003}. It is clear that if $B = \{ \mathbf{0} \}$, then $\epsilon_B = \delta_B = \id_{\mathcal{I}_P}$ where $\id_{\mathcal{I}_P} : \mathcal{I}_P \rightarrow \mathcal{I}_P$ denotes the identity function and $O_B = C_B = \id_{\mathcal{I}_P}$.

We conclude this section by reviewing some basic properties of these morphological operations.  We will use these properties to establish our main result. 

\begin{proposition} 
[Properties 3.4~\cite{Soille2003}, p. 71]
\label{prop. dilation and erosion are increasing}
Let $f, g \in \mathcal{I}_P$ be images and $B \subseteq \mathbb{Z}^m$ be a structuring element. If $f \leq g$, then the following inequalities hold
\begin{equation*}
    \delta_B(f) \leq \delta_B(g),~ \epsilon_B(f) \leq \epsilon_B(g),~ O_B(f) \leq O_B(g),~\text{and } C_B(f) \leq C_B(g).
\end{equation*}
\end{proposition}

Proposition \ref{prop. dilation and erosion are increasing} states the {\em increasing property}, that is, for a fixed structuring element, the basic morphological operations preserve the ordering relation on images.

In Definition \ref{def:image}, we define images as functions.  Our main focus in this work is to construct a {\em filtration} or collection of sets ordered by set inclusion.  We do this for image sublevel sets, i.e. subsets of the pixel set $P$ corresponding to pixels with image values at or below a prescribed threshold value. When we consider binary images, the filtration property of sublevel sets is naturally related to the increasing property as shown in the following proposition.

\begin{proposition}
[Principle 11.1.1~\cite{Soille2003}, p. 318]
\label{Prop. Equivalent property of g <= f}
Let $f, g \in \mathcal{I}_P$ be images. If $f \leq g$, then $g^{-1}(0) \subseteq f^{-1}(0)$. In addition, if $f, g : P \rightarrow \{ 0,1 \}$ are binary images, then $f \leq g$ if and only if $g^{-1}(0) \subseteq f^{-1}(0)$ and, similarly, $f \leq g$ if and only if $f^{-1}(1) \subseteq g^{-1}(1)$. 
\end{proposition}

There are many ways to produce a binary image from a grayscale image.  {\em Global thresholding} of grayscale image $g:P\rightarrow \mathbb{R}_{\geq0}$ via threshold value $t$ produces the binary image 
\begin{equation}
\label{Equation : Grayscale to binary with threshold t}
 	g_{t}(\bfx) = \begin{cases}
 	0 \quad\null\text{ if } g(\bfx) \leq t, \\
 	1 \quad\null\text{ otherwise.}
 	\end{cases}
\end{equation}

Note that $g_t^{-1}(0)=\{\bfx\in P:g(\bfx)\leq t\}$.  This set, $g_t^{-1}(0)$ is the {\em $t$-sublevel set} of $g$.
In general, the operations of erosion, dilation, opening, and closing do not commute.  However, these four operations do commute with the operation of global thresholding as follows.
\begin{proposition}
[Proposition 1, \cite{hu2020conditions}]
\label{Proposition : Ristriction and opening/closing are commutative}
For $m \in \mathbb{N}$ and pixel set $P \subseteq \bbZ^m$, consider the image $g \in \mathcal{I}_P$. For each threshold $t \in \mathbb{R}_{\geq 0}$ we define $\tau_t : \mathcal{I}_P \rightarrow \mathcal{I}_P$ by $g \mapsto g_t$ i.e.,  
\begin{equation}
 	\tau_t(g)(\bfx) = g_{t}(\bfx) = \begin{cases}
 	0 \quad\null\text{ if } g(\bfx) \leq t, \\
 	1 \quad\null\text{ otherwise.}
 	\end{cases}.
\end{equation}
For any structuring element $B \subseteq \mathbb{Z}^m$, the following diagrams are commutative:
\begin{equation*}
\begin{split}
\xymatrix@+1.0em{
				& \mathcal{I}_P
				\ar[d]_{\epsilon_B}
				\ar[r]^{\tau_t}
                & \mathcal{I}_P
                \ar[d]^{\epsilon_B}
        		\\
        		& \mathcal{I}_P
				\ar[r]^{\tau_t}
                & \mathcal{I}_P
}
\ \
\xymatrix@+1.0em{
				& \mathcal{I}_P
				\ar[d]_{\delta_B}
				\ar[r]^{\tau_t}
                & \mathcal{I}_P
                \ar[d]^{\delta_B}
        		\\
        		& \mathcal{I}_P
				\ar[r]^{\tau_t}
                & \mathcal{I}_P
}
\end{split}
\end{equation*}
i.e., $\epsilon_B \circ \tau_t = \tau_t \circ \epsilon_B$ and $\delta_B \circ \tau_t = \tau_t \circ \delta_B$.  Moreover, by combining these two commutative diagrams, $O_B \circ \tau_t = \tau_t \circ O_B \ \ {and} \ \ C_B \circ \tau_t = \tau_t \circ C_B.$
\end{proposition}

Finally we note the relationship between the partial order on grayscale images and the partial order on the corresponding thresholded images.

\begin{lemma}
[Lemma 1, \cite{hu2020conditions}]
\label{Lemma : Threshold inclusion relation}
For images $f, g \in \mathcal{I}_P$, $f \leq g$ if and only if $f_t \leq g_t$ for every $t \in \bbR_{\geq 0}$.
\end{lemma}

Finally, dilation and erosion using structuring elements related by inclusion also preserve the ordering relation.
\begin{proposition}
[\cite{Soille2003}]
\label{Prop:erosion dilation subset}
%\label{Proposition : fixed image, different B, and their relations}
Let $B_1 \subseteq B_2 \subseteq \mathbb{Z}^m$ be structuring elements.  Then for $g \in \mathcal{I}_P$,
%For image $g \in \mathcal{I}_P$ and structuring elements $B_1 \subseteq B_2 \subseteq \mathbb{Z}^m$, 
\begin{equation*}
    \delta_{B_1}(g) \leq \delta_{B_2}(g){\text{, and }} \epsilon_{B_2}(g) \leq \epsilon_{B_1}(g).
\end{equation*}
\end{proposition}

\section{One-Parameter Filtrations and Persistent Homology}
\label{sec:ph}
In this section, we show that the partial order results for morphological operations presented in Section \ref{sec:background} naturally yield the structure necessary for computing {\it persistent homology}.  

Persistent homology, a foundational tool in the field of topological data analysis (TDA), measures and tracks topological features.  It relies on having a {\it one-parameter filtration}, a sequence of nested sets.  The goal of this section is to introduce two new filtrations based on morphological operations and define and illustrate the meaning of {\it persistence diagrams} based on these filtrations.

Topological features of interest include connected components (or individual connected pieces of the set), one-dimensional holes (holes in 2d or tunnels in 3d) and two-dimensional holes (cavities in 3d).  Higher dimensional holes may appear in higher dimensional data, but for illustration, we will focus on two and three dimensional data sets here.  The framework we present throughout this work applies to higher dimensional data and holes as well.  For the data sets we study, cubical homology may be summarized using Betti numbers.  Betti numbers, $\beta_k$, count holes of various dimensions.  More specifically, given a binary image $f$, if we consider the set of black pixels, $X:=f^{-1}(0)$, then $\beta_0(X)$ is the number of connected components, $\beta_1(X)$ is the number of 1-dimensional holes, or tunnels, $\beta_2(X)$ is the number of 2-dimensional holes, or cavities, etc.  They are computed using algebraic structure defined by the cubical structure of $X$ and there are now efficient software packages for performing these calculations.  See, for example, \cite{kaczynski2004computational} and references therein for a discussion of the mathematical theory behind the definition and computation of Betti numbers as well as their interpretation as direct counts of topological features. 

Persistent homology extends the topological measurement offered by Betti numbers across a filtration. A {\em one-parameter filtration} is a sequence of sets $\{X_i\}_{i\in A}$, with indexing set $A\subset \mathbb{Z}$, satisfying
\begin{equation}
\label{equ:def of one filtration}
    X_i \subseteq X_j, \quad\null \text{whenever } i\leq j.
\end{equation}

For ease of notation, we will often write $\{X_i\}$ when the indexing set has already been specified.  For a one-parameter filtration, persistent homology records the {\it birth} and {\it death} coordinates at which a given topological feature first appears and first disappears respectively in cubical sets. That is, given a one-parameter filtration $\{X_i\}_{i\in A}$ of cubical sets, $X_i$, a feature with birth/death coordinates $(b,d)$, does not exist in the sets $X_i$ with $i<b$, appears first in $X_b$ and persists through all sets $X_j$ with $b\leq j < d$ and disappears in $X_{d}$. Like Betti numbers, birth/death coordinates are computed using algebraic structure attached to cubical sets, in this case using the inclusion operator to match some features in $X_n$ to their preimages in $X_{n-1}$.

The collection of birth/death pairs for all topological features, labeled by dimension of the feature, is called a {\it persistence diagram}. For a given one-parameter filtration $\{X_i\}$, the full persistence diagram is the collection of all birth/death pairs and is denoted by ${\cal{P}}(\{X_i\})$, with the {\it k-th persistence diagram}, ${\cal{P}}_k(\{X_i\})$, denoting the subset of pairs measuring $k$-dimensional holes.  By this convention, ${\cal{P}}(\{X_i\}) = \displaystyle\cup_{k}{\cal{P}}_k(\{X_i\})$.  For ease of notation, we represent these persistence diagrams by ${\cal{P}}$ (respectively ${\cal{P}}_k$) when the filtration $\{X_i\}$ is understood. Betti numbers may be extracted from persistence diagrams as
\begin{equation}\label{eqn:betti_count}
	\beta_k( X_m ) = \#{\cal{P}}_k(m),
\end{equation}
where
\begin{equation}
	{\cal{P}}_k(m) := \{(b,d)\in {\cal{P}}_k \mid\; b\leq m,\; d>m\}.
\end{equation}

In other words, $\beta_k( X_m )$ counts the number of points in the k-th persistence diagram whose birth/death coordinates indicate that they are present in set $X_m$. By extension, we also write ${\cal{P}}(m):=\cup_k {\cal{P}}_k(m)$.  Furthermore, a given feature's lifespan, $l=d-b$, measures the length of the interval of set indices over which the feature persists.  Birth/death coordinates and corresponding lifespans allow us to study the robustness of the feature with respect to changes in the index $m$.

For more information about persistent homology,  see e.g. \cite{zomorodian2005computing,edelsbrunner2008persistent},  and references therein. In summary, given a one-parameter filtration of cubical sets, persistence diagrams are efficient to compute.  The reference \cite{otter2017roadmap} provides an overview of current TDA software.  In this work, we use {\tt Perseus} \cite{perseus} and {\tt DIPHA} \cite{DIPHA} for cubical persistent homology computations.

Researchers have developed several methods for creating one-parameter filtrations. The most fundamental and commonly-used filtration for grayscale digital images is the {\em sublevel set filtration} (see, e.g. \cite{GarinAndTauzin2019,BernsteinEtAL2020}).  Using the sublevel set and thresholding operations \eqref{Equation : Grayscale to binary with threshold t}, for thresholds $t_1 \leq t_2 \leq \cdots \leq t_n$,

\begin{equation}
\label{equ:sublevel set filtration}
     g^{-1}_{t_1}(0) \subseteq g^{-1}_{t_2}(0) \subseteq \cdots g^{-1}_{t_n}(0).
\end{equation}

Setting $X_i=g^{-1}_{t_i}(0)$ yields a sublevel set filtration $\{X_i\}$.   For binary images, there are techniques for constructing related grayscale images that would then lead to sublevel set filtrations.  These include, for example, using a signed distance function or density estimator to define grayscale values \cite{obayashi2018persistence,garin2019topological}.   {While \cite{FrosiniAndLandi1997} considers the changes of size functions (the 0-th persistence diagram of the sublevel set filtration) under the skeleton operation which combines certain morphological operations, our goal here is build a general filtration framework using erosion, dilation, opening, and closing. To the best of our knowledge, the proposed work is the first to use morphological operations and thresholding to construct a filtration directly. }

We now use morphological operations to form new filtrations for binary and grayscale images.  For ease of discussion, throughout the article we use a sequence of structuring elements, $\{B_i\}_{i=0}^n$, where each $B_i$ is a $(i+1)\times (i+1)$ square given by
\begin{equation}
    \label{eqn:1, 4, 9, 16,... rectangles interation formula}
    \begin{split}
    B_0 &= \{ (0,0) \},\\ 
    B_n &= \left\{ \begin{array}{ll}
B_{n-1} \cup (B_{n-1} + \bfe_1) \cup (B_{n-1} + \bfe_2) \cup (B_{n-1} + \bfe_1 + \bfe_2) & \mbox{if $n$ is odd}\\
B_{n-1} \cup (B_{n-1} - \bfe_1) \cup (B_{n-1} - \bfe_2) \cup (B_{n-1} - \bfe_1 - \bfe_2) & \mbox{if $2n$ is even, $n \geq 2$}
\end{array}\right. ,    
    \end{split}
\end{equation}
where $\bfe_1 = (1,0)$ and $\bfe_2=(0, 1)$. These may be depicted as
\begin{equation}
\label{eqn:1, 4, 9, 16,... rectangles}
B_0 = 
    \begin{array}{c}
\circ \\
\end{array} \ \ ,  \ \
B_1 = 
    \begin{array}{cc}
\bullet & \bullet \\ 
\circ & \bullet \\
\end{array} \ \ ,  \ \
B_2 = 
    \begin{array}{ccc}
\bullet & \bullet & \bullet \\
\bullet & \circ & \bullet \\ 
\bullet & \bullet & \bullet \\
\end{array} \ \ ,  \ \
B_3 = 
    \begin{array}{cccc}
\bullet & \bullet & \bullet & \bullet \\
\bullet & \bullet & \bullet & \bullet \\ 
\bullet & \circ & \bullet & \bullet \\
\bullet & \bullet & \bullet & \bullet
\end{array}  \ \ ,~\cdots \ \
\end{equation}
where $\circ$ represents the origin $(0,0) \in \mathbb{Z}^2$.  Clearly, $B_0\subseteq B_1 \subseteq \cdots \subseteq B _n$.   Note that since $B_0=\{(0,0)\}$, the erosion/dilation, and opening/closing operations with respect to $B_0$ are the identity map.  
Other sequences of structuring elements will also give rise to filtrations.  In particular, there is a notion of {\em shift inclusion} that may be used to designate a large class of sequences of structuring elements that may be used to form filtrations.  That topic is studied in detail in \cite{hu2020conditions}.

The first new filtrations we propose are for binary images and use the erosion and dilation operations.  We consider a sequence of erosion and dilation operations with respect to $\{B_i\}_{i=0}^n$, i.e. for each $i$, consider $\delta_{B_i}(f)$ and $\epsilon_{B_i}(f)$ for a given binary image $f$.  Similar to the sublevel set filtration in \eqref{equ:sublevel set filtration}, the desired property is that if $i\leq j$ ($B_i \subseteq B_j$), then $\epsilon_{B_i}(f)^{-1}(0) \subseteq \epsilon_{B_j}(f)^{-1}(0) $. Thanks to Proposition \ref{Prop. Equivalent property of g <= f} and Proposition \ref{Prop:erosion dilation subset}, it is straightforward to verify that
\begin{align}
& \epsilon_{B_0}(f)^{-1}(0) \subseteq \epsilon_{B_1}(f)^{-1}(0) \subseteq \cdots \subseteq \epsilon_{B_n}(f)^{-1}(0),\label{equ:erosion filtration}\\
    & \delta_{B_n}(f)^{-1}(0)  \subseteq \delta_{B_{n-1}}(f)^{-1}(0) \subseteq \cdots \subseteq \delta_{B_0}(f)^{-1}(0). \label{equ:dilation filtration}
\end{align}

This shows that for any sequence of nested structural elements, erosion and dilation form filtrations. {We call $\{X^{\epsilon}_i\}_{i=0}^n$, where $X^{\epsilon}_i=\epsilon_{B_i}(f)^{-1}(0)$, the {\it erosion filtration}, and $\{X^{\delta}_j\}_{j=-n}^0$, where $X^{\delta}_j=\delta_{B_{|j|}}(f)^{-1}(0)$, the {\it dilation filtration}.  Note that since $\epsilon_{B_0}(f)^{-1}(0) = f^{-1}(0) = \delta_{B_0}(f)^{-1}(0)$, we may form one extended filtration by taking $\{\tilde{X}_i\}_{i=-n}^n$, where for $i<0$, $\tilde{X}_i= X^{\delta}_i$, and for $i>0$, $\tilde{X}_i= X^{\epsilon}_i$. }

The second new filtration we propose is related to the opening and closing operations.  Since opening and closing are compositions of erosion and dilation operations, one may expect that Proposition \ref{Prop:erosion dilation subset} would extend to the case of opening or closing.  However, it is not true in general.  We refer readers to \cite{hu2020conditions} for a counter example and more discussion.  Essentially, the sequence of structuring elements cannot be arbitrary and requires additional assumptions. \cite{hu2020conditions} presents a sufficient condition called {\em shift inclusion} that guarantees the structure necessary for opening and closing to result in appropriately nested sets.

Since our chosen square structuring elements, $B_i$, satisfy shift inclusion \cite{hu2020conditions},  $O_{B_i}$ and, separately, $C_{B_i}$, also form filtrations.

\begin{align}
& O_{B_0}(f)^{-1}(0) \subseteq O_{B_1}(f)^{-1}(0) \subseteq \cdots \subseteq O_{B_n}(f)^{-1}(0);\label{equ:opening filtration}\\
    & C_{B_n}(f)^{-1}(0)  \subseteq C_{B_{n-1}}(f)^{-1}(0) \subseteq \cdots \subseteq C_{B_0}(f)^{-1}(0). \label{equ:closing filtration}
\end{align}

Similar to erosion and dilation filtration, we call $\{X^{\mathcal{O}}_i\}_{i=0}^n$, where $X^{\mathcal{O}}_i= O_{B_i}(f)^{-1}(0)$, the {\it opening filtration}, and $\{X^{\mathcal{C}}_j\}_{j=-n}^0$, where $X^{\mathcal{C}}_j=C_{B_{|j|}}(f)^{-1}(0)$, the {\it closing filtration}.  Note that since $O_{B_0}(f)^{-1}(0) = f^{-1}(0) = C_{B_0}(f)^{-1}(0)$, we may form one extended filtration by taking $\{\tilde{X}_i\}_{i=-n}^n$, where for $i<0$, $\tilde{X}_i= X^{\mathcal{C}}_i$, and for $i>0$, $\tilde{X}_i= X^{\mathcal{O}}_i$.

As a byproduct, applications of \eqref{equ:opening filtration} and \eqref{equ:closing filtration} lead to three additional filtrations based on the commonly used top-hat transformation: the \textit{white top hat} $\text{WTH}_B(f) = f - O_B(f)$, the \textit{black top hat} $\text{BTH}_B(f) = C_B(f) - f$, and the \textit{self complementary top hat transformation} $\text{STH}_B(f) = C_B(f) - O_B(f)$~\cite{dougherty1992introduction,serra1984image,Soille2003}. More precisely, one has
\begin{align}
    & \text{WTH}_{B_n}(f)^{-1}(0)  \subseteq \text{WTH}_{B_{n-1}}(f)^{-1}(0) \subseteq \cdots \subseteq \text{WTH}_{B_0}(f)^{-1}(0), \label{equ:wth filtration}\\
    & \text{BTH}_{B_n}(f)^{-1}(0)  \subseteq \text{BTH}_{B_{n-1}}(f)^{-1}(0) \subseteq \cdots \subseteq \text{BTH}_{B_0}(f)^{-1}(0), \label{equ:bth filtration}\\
    & \text{STH}_{B_n}(f)^{-1}(0)  \subseteq \text{STH}_{B_{n-1}}(f)^{-1}(0) \subseteq \cdots \subseteq \text{STH}_{B_0}(f)^{-1}(0). \label{equ:sth filtration}
\end{align}

\begin{figure}
	\centering
%	\subfloat[$X$]{\includegraphics[width=0.4\linewidth]{kanjimorO1.png}}~
 	\subfloat[$X^{\mathcal{O}}_0$]{\includegraphics[width=0.19\linewidth]{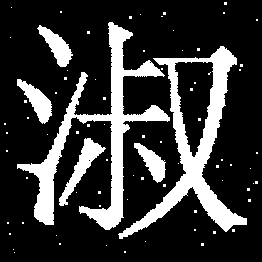}}~
 	\subfloat[$X^{\mathcal{O}}_1$]{\includegraphics[width=0.19\linewidth]{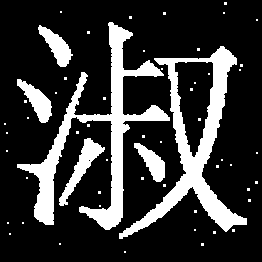}}~
 %	\subfloat[$X_2$]{\includegraphics[width=0.19\linewidth]{kanjiM_1_3.png}}~
 	\subfloat[$X^{\mathcal{O}}_3$]{\includegraphics[width=0.19\linewidth]{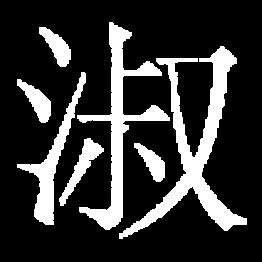}}~
 	\subfloat[$X^{\mathcal{O}}_5$]{\includegraphics[width=0.19\linewidth]{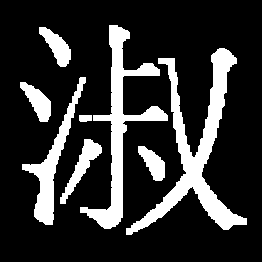}}~ 	
 	\subfloat[$X^{\mathcal{O}}_7$]{\includegraphics[width=0.19\linewidth]{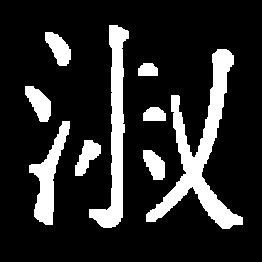}}\\
 	\subfloat[$X^{\mathcal{O}}_{9}$]{\includegraphics[width=0.19\linewidth]{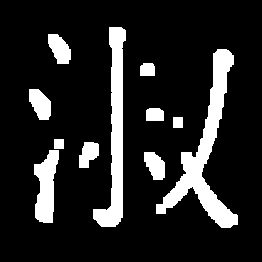}}~
 	\subfloat[$X^{\mathcal{O}}_{11}$]{\includegraphics[width=0.19\linewidth]{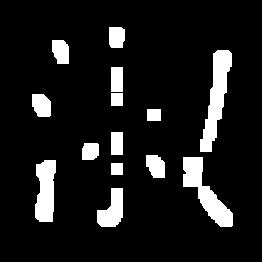}}~ 	
 	\subfloat[$X^{\mathcal{O}}_{13}$]{\includegraphics[width=0.19\linewidth]{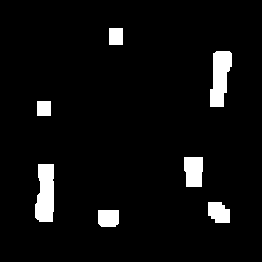}}~
 	\subfloat[$X^{\mathcal{O}}_{15}$]{\includegraphics[width=0.19\linewidth]{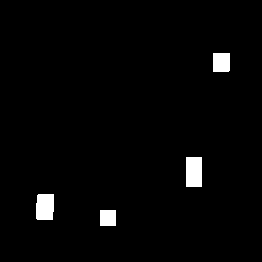}}~
 	\subfloat[$X^{\mathcal{O}}_{17}$]{\includegraphics[width=0.19\linewidth]{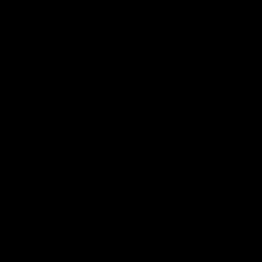}}\\
	\subfloat[$\mathcal{P}_1{[}\{X^{\mathcal{O}}_i\}_i{]}$]{\includegraphics[width=0.55\linewidth]{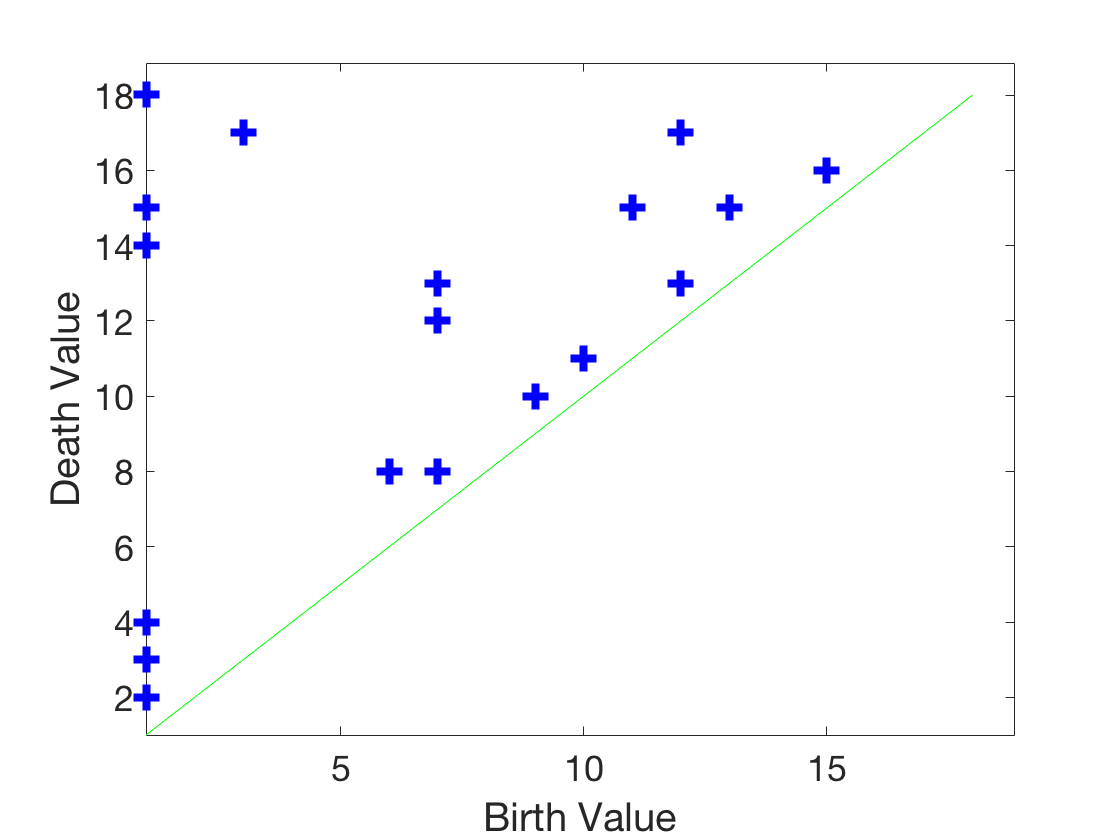}}~
	\caption{Opening filtration and its persistence diagram.  (a) The original binary image, $f$, is Kanji with salt noise; (b)-(j) binary representations of $X^{\mathcal{O}}_i := O(f)^{-1}_{B_i}(0)$ with $i=0, 1, 3, \ldots, 17$.; (k) The 1-st persistence diagram $\mathcal{P}_1[\{X^{\mathcal{O}}_i\}_{i=0}^{18}]$ for the opening filtration $X^{\mathcal{O}}_i=O(f)^{-1}_{B_i}(0)$.}	
	\label{fig:conceptual}
\end{figure}

For illustration, we now present a relatively simple opening filtration.  Consider the modified Kanji image shown in Figure \ref{fig:conceptual}. The original binary image is shown in Figure \ref{fig:conceptual}(a) and denoted by $f$.  Sample sets from the opening filtration on $f$, $X^{\mathcal{O}}_i$, $i=0,\ldots, 18$, are shown in the top two rows of Figure \ref{fig:conceptualbiFiltration} and the corresponding 1st level persistence diagram is shown in Figure \ref{fig:conceptualbiFiltration}(k).  By \eqref{eqn:betti_count}, we know that $\beta_1(X^{\mathcal{O}}_0) = \# \mathcal{P}_1(0)$.  In particular, $\mathcal{P}_1(0)$ consists of the points on the vertical axis of Figure \ref{fig:conceptual}(k).

By construction of the opening filtration, features in the original image {$0$-level} set, $X^{\mathcal{O}}_0$, (having birth coordinate $b=0$), that are small in spatial scale relative to the structuring elements, have a short lifespan (small death coordinate $d$) whereas larger scale features have a longer lifespan (large $d$).  The separation in scale between spatially small and large features is evident on the left vertical axis ($b=0$) in the persistence diagram in Figure \ref{fig:conceptual}(f).   

Features in $X^{\mathcal{O}}_0$ that disappear quickly under small amounts of opening (that is features with $b=0$ and small $d$), have a small geometric/spatial scale, features in $X_0$ that persist under a lot of opening ($b=0$ and large $d$) are more robust, and features that only appear after a lot of opening (large $b$) are most likely spurious.  Choosing $m=4$ in this example allows us to separate these three groups by drawing horizontal and vertical lines at $d=4$ and $b=4$ respectively.  Correspondingly, $X^{\mathcal{O}}_4$ contains only so-called robust features and $\beta_1(X^{\mathcal{O}}_4)$ is a count of these features. One of the captured features, however, has a birth coordinate $b>0$ indicating that it was not present in the original image.

We now return to the more general erosion/dilation and opening/closing one-parameter filtrations presented earlier and extend these to form filtrations on grayscale images.  Combining either of these filtrations with the sublevel set filtration in \eqref{equ:sublevel set filtration}, one may obtain a {\it two-parameter filtration, or bi-filtration}. We take the opening filtration as an illustration.  Given a grayscale image $g$, by \eqref{equ:sublevel set filtration}, we have $g^{-1}_{t_i}(0) \subseteq g^{-1}_{t_j}(0)$ for any $t_i \leq t_j$.  Since for each $t$, $g_t$ is a binary image, by \eqref{equ:opening filtration} we have that $O_{B_i}(g_t)^{-1}(0) \subseteq O_{B_j}(g_t)^{-1}(0)$.  By combining both \eqref{equ:sublevel set filtration} and \eqref{equ:opening filtration} we obtain
{\small
\begin{equation}
	\label{equ:bifiltration O and t}
	\begin{tabular}{ccccccc}
		$O_{B_N}(g_{t_1})^{-1}(0)$ & $\subseteq$ & $O_{B_N}(g_{t_2})^{-1}(0)$ & $\subseteq\cdots\subseteq$ & $O_{B_N}(g_{t_{T-1}})^{-1}(0)$ & $\subseteq$ & $O_{B_N}(g_{t_T})^{-1}(0)$  \\
		\rotatebox[origin=c]{90}{$\subseteq$} &  & \rotatebox[origin=c]{90}{$\subseteq$} & $\vdots$& \rotatebox[origin=c]{90}{$\subseteq$} &  & \rotatebox[origin=c]{90}{$\subseteq$} \\	
		$\vdots$ &  & $\vdots$ & $\vdots$& $\vdots$&& $\vdots$ \\			
		\rotatebox[origin=c]{90}{$\subseteq$} &  & \rotatebox[origin=c]{90}{$\subseteq$} & $\vdots$& \rotatebox[origin=c]{90}{$\subseteq$} &  & \rotatebox[origin=c]{90}{$\subseteq$} \\	
		$O_{B_2}(g_{t_1})^{-1}(0)$ & $\subseteq$ & $O_{B_2}(g_{t_2})^{-1}(0)$ & $\subseteq\cdots\subseteq$ & $O_{B_2}(g_{t_{T-1}})^{-1}(0)$ & $\subseteq$ & $O_{B_2}(g_{t_T})^{-1}(0)$ \\
		\rotatebox[origin=c]{90}{$\subseteq$} &  & \rotatebox[origin=c]{90}{$\subseteq$} & $\vdots$& \rotatebox[origin=c]{90}{$\subseteq$} &  & \rotatebox[origin=c]{90}{$\subseteq$} \\	
		$O_{B_1}(g_{t_1})^{-1}(0)$ & $\subseteq$ & $O_{B_1}(g_{t_2})^{-1}(0)$ & $\subseteq\cdots\subseteq$ & $O_{B_1}(g_{t_{T-1}})^{-1}(0)$ & $\subseteq$ & $O_{B_1}(g_{t_T})^{-1}(0)$  \\
		\rotatebox[origin=c]{90}{$\subseteq$} &  & \rotatebox[origin=c]{90}{$\subseteq$} & $\vdots$& \rotatebox[origin=c]{90}{$\subseteq$} &  & \rotatebox[origin=c]{90}{$\subseteq$} \\	
		$g^{-1}_{t_1}(0)$ & $\subseteq$ & $g^{-1}_{t_2}(0)$ & $\subseteq\cdots\subseteq$ & $g^{-1}_{t_{T-1}}(0)$ & $\subseteq$ & $g^{-1}_{t_T}(0)$
	\end{tabular}.
\end{equation}
}

This is a $2$-filtration, an example of a multi-filtration defined in Definition~\ref{def:multi filtration}.  In the next section, we formalize and extend the class of multi-filtrations constructed from opening and closing operations on binary images and opening, closing, and thresholding operations for grayscale images.

\section{Multi-parameter Filtrations}
\label{sec:multiparameter filtrations}
At this point, we have seen that erosion, dilation, opening, and closing each form one-parameter filtrations for binary images and that combining one of these operations with thresholding forms a $2$-filtration (see Definition~\ref{def:multi filtration} below).  As we show in the next example, opening and closing operations may also be combined to form a $2$-filtration for a binary image.  In fact, this process may be continued to define $k$-parameter filtrations, the overall goal of this section.

\begin{definition}[\cite{carlsson2010computing}]
\label{def:multi filtration}
For $k \in \bbN$ and $\mathbf{u},~\mathbf{v}\in \mathbb{Z}^k$ we say that $\mathbf{u} \leq \mathbf{v}$ if and only if $u_i \leq v_i$ for all $i$. Given this partial order on $\mathbb{Z}^k$, a family of sets $\{S_{\mathbf{i}}\}_{{\mathbf{i}}\in A}$ with indexing set $A \subseteq \mathbb{Z}^k$ is a \textit{multifiltration} (or \textit{$k$-parameter filtration}) if for any $\mathbf{u}, \mathbf{v}\in A$ with $\mathbf{u}\leq \mathbf{v}$, $S_{\mathbf{u}} \subseteq S_{\mathbf{v}}$.  
\end{definition}

Combining an opening filtration and a closing filtration and invoking Proposition \ref{prop. dilation and erosion are increasing} yields the following $2$-filtration.
{\small
\begin{equation}
	\label{equ:bifiltration O and C}
	\begin{tabular}{ccccccc}
		$f^{-1}(0)$ & $\subseteq$ & $O_{B_2}(f)^{-1}(0) $ & $\subseteq\cdots\subseteq$ & $O_{B_{N-1}}(f)^{-1}(0) $ & $\subseteq$ & $O_{B_N}(f)^{-1}(0) $  \\
		\rotatebox[origin=c]{90}{$\subseteq$} &  & \rotatebox[origin=c]{90}{$\subseteq$} & $\vdots$& \rotatebox[origin=c]{90}{$\subseteq$} &  & \rotatebox[origin=c]{90}{$\subseteq$} \\	
		$C_{B_2}(f)^{-1}(0)$ & $\subseteq$ & $C_{B_2}(O_{B_2}(f))^{-1}(0) $ & $\subseteq\cdots\subseteq$ & $C_{B_2}(O_{B_{N-1}}(f))^{-1}(0) $ & $\subseteq$ & $C_{B_2}(O_{B_N}(f))^{-1}(0) $  \\		
		\rotatebox[origin=c]{90}{$\subseteq$} &  & \rotatebox[origin=c]{90}{$\subseteq$} & $\vdots$& \rotatebox[origin=c]{90}{$\subseteq$} &  & \rotatebox[origin=c]{90}{$\subseteq$} \\			
		$C_{B_3}(f)^{-1}(0)$ & $\subseteq$ & $C_{B_3}(O_{B_2}(f))^{-1}(0) $ & $\subseteq\cdots\subseteq$ & $C_{B_3}(O_{B_{N-1}}(f))^{-1}(0) $ & $\subseteq$ & $C_{B_3}(O_{B_N}(f))^{-1}(0) $  \\		
		\rotatebox[origin=c]{90}{$\subseteq$} &  & \rotatebox[origin=c]{90}{$\subseteq$} & $\vdots$& \rotatebox[origin=c]{90}{$\subseteq$} &  & \rotatebox[origin=c]{90}{$\subseteq$} \\					
		$\vdots$ &  & $\vdots$ & $\vdots$& $\vdots$&& $\vdots$ \\			
		\rotatebox[origin=c]{90}{$\subseteq$} &  & \rotatebox[origin=c]{90}{$\subseteq$} & $\vdots$& \rotatebox[origin=c]{90}{$\subseteq$} &  & \rotatebox[origin=c]{90}{$\subseteq$} \\
		$C_{B_N}(f)^{-1}(0)$ & $\subseteq$ & $C_{B_N}(O_{B_2}(f))^{-1}(0) $ & $\subseteq\cdots\subseteq$ & $C_{B_N}(O_{B_{N-1}}(f))^{-1}(0) $ & $\subseteq$ & $C_{B_N}(O_{B_N}(f))^{-1}(0) $ 
	\end{tabular}.
\end{equation}
}

\begin{figure}
	\centering
	\captionsetup[subfigure]{labelformat=empty}
	\subfloat[$X^{\mathcal{O},\mathcal{C}}_{(0,0)}$]{\includegraphics[width=0.19\linewidth]{kanjiM_1_1.png}}~
%	\subfloat[$X_{(0,1)}$]{\includegraphics[width=0.15\linewidth]{kanjiM_1_2.png}}~
	\subfloat[$X^{\mathcal{O},\mathcal{C}}_{(0,2)}$]{\includegraphics[width=0.19\linewidth]{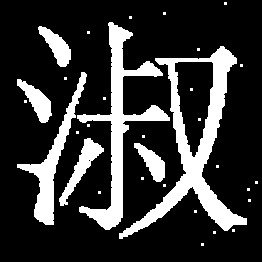}}~
	\subfloat[$X^{\mathcal{O},\mathcal{C}}_{(0,3)}$]{\includegraphics[width=0.19\linewidth]{kanjiM_1_4.png}}~
	\subfloat[$X^{\mathcal{O},\mathcal{C}}_{(0,9)}$]{\includegraphics[width=0.19\linewidth]{kanjiM_1_10.png}}~
	\subfloat[$X^{\mathcal{O},\mathcal{C}}_{(0,13)}$]{\includegraphics[width=0.19\linewidth]{kanjiM_1_14.png}}\\

\subfloat[$X^{\mathcal{O},\mathcal{C}}_{(-1,0)}$]{\includegraphics[width=0.19\linewidth]{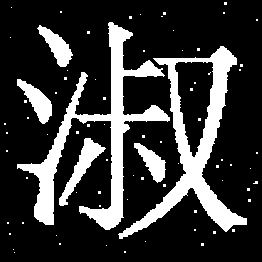}}~
%\subfloat[$X_{(1,1)}$]{\includegraphics[width=0.15\linewidth]{kanjiM_2_2.png}}~
\subfloat[$X^{\mathcal{O},\mathcal{C}}_{(-1,2)}$]{\includegraphics[width=0.19\linewidth]{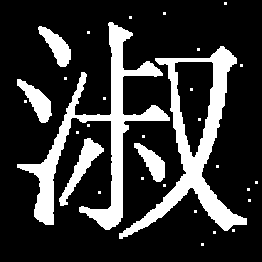}}~
\subfloat[$X^{\mathcal{O},\mathcal{C}}_{(-1,3)}$]{\includegraphics[width=0.19\linewidth]{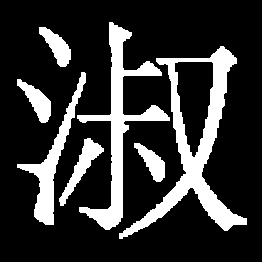}}~
\subfloat[$X^{\mathcal{O},\mathcal{C}}_{(-1,9)}$]{\includegraphics[width=0.19\linewidth]{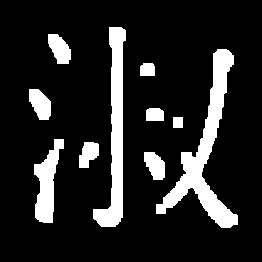}}~
\subfloat[$X^{\mathcal{O},\mathcal{C}}_{(-1,13)}$]{\includegraphics[width=0.19\linewidth]{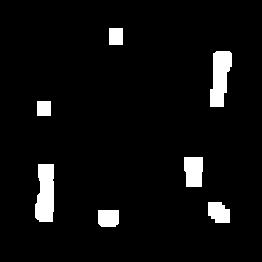}}\\

\subfloat[$X^{\mathcal{O},\mathcal{C}}_{(-2,0)}$]{\includegraphics[width=0.19\linewidth]{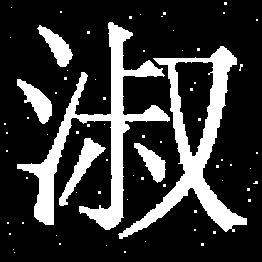}}~
%\subfloat[$X_{(2,1)}$]{\includegraphics[width=0.15\linewidth]{kanjiM_3_2.png}}~
\subfloat[$X^{\mathcal{O},\mathcal{C}}_{(-2,2)}$]{\includegraphics[width=0.19\linewidth]{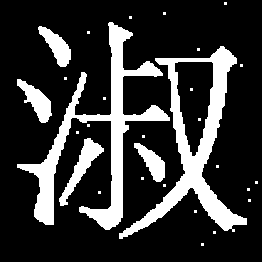}}~
\subfloat[$X^{\mathcal{O},\mathcal{C}}_{(-2,3)}$]{\includegraphics[width=0.19\linewidth]{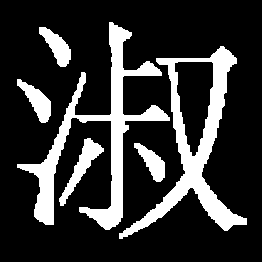}}~
\subfloat[$X^{\mathcal{O},\mathcal{C}}_{(-2,9)}$]{\includegraphics[width=0.19\linewidth]{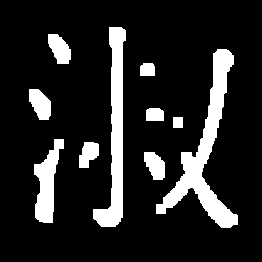}}~
\subfloat[$X^{\mathcal{O},\mathcal{C}}_{(-2,13)}$]{\includegraphics[width=0.19\linewidth]{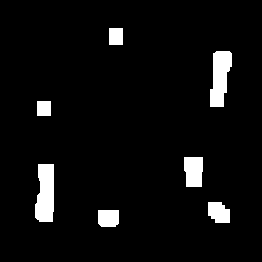}}\\

\subfloat[$X^{\mathcal{O},\mathcal{C}}_{(-8,0)}$]{\includegraphics[width=0.19\linewidth]{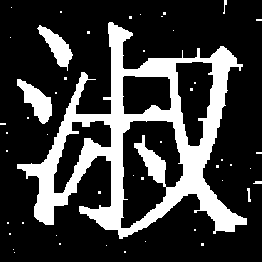}}~
%\subfloat[$X_{(8,1)}$]{\includegraphics[width=0.15\linewidth]{kanjiM_9_2.png}}~
\subfloat[$X^{\mathcal{O},\mathcal{C}}_{(-8,2)}$]{\includegraphics[width=0.19\linewidth]{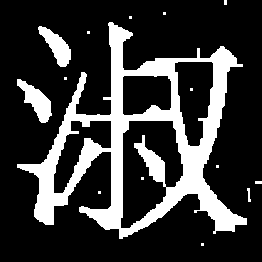}}~
\subfloat[$X^{\mathcal{O},\mathcal{C}}_{(-8,3)}$]{\includegraphics[width=0.19\linewidth]{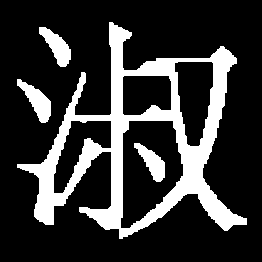}}~
\subfloat[$X^{\mathcal{O},\mathcal{C}}_{(-8,9)}$]{\includegraphics[width=0.19\linewidth]{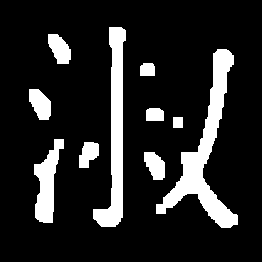}}~
\subfloat[$X^{\mathcal{O},\mathcal{C}}_{(-8,13)}$]{\includegraphics[width=0.19\linewidth]{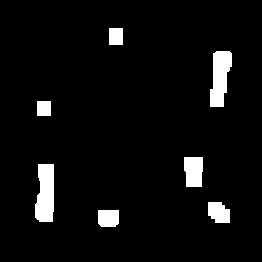}}\\

\subfloat[$X^{\mathcal{O},\mathcal{C}}_{(-16,0)}$]{\includegraphics[width=0.19\linewidth]{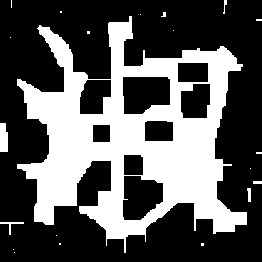}}~
%\subfloat[$X_{(16,1)}$]{\includegraphics[width=0.15\linewidth]{kanjiM_17_2.png}}~
\subfloat[$X^{\mathcal{O},\mathcal{C}}_{(-16,2)}$]{\includegraphics[width=0.19\linewidth]{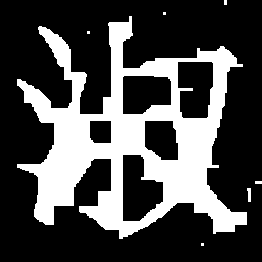}}~
\subfloat[$X^{\mathcal{O},\mathcal{C}}_{(-16,3)}$]{\includegraphics[width=0.19\linewidth]{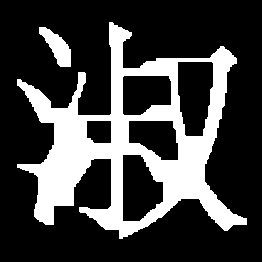}}~
\subfloat[$X^{\mathcal{O},\mathcal{C}}_{(-16,9)}$]{\includegraphics[width=0.19\linewidth]{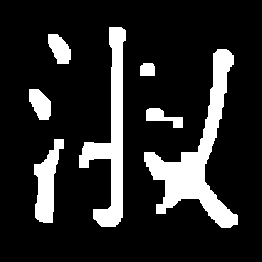}}~
\subfloat[$X^{\mathcal{O},\mathcal{C}}_{(-16, 13)}$]{\includegraphics[width=0.19\linewidth]{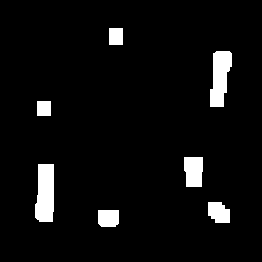}}
	\caption{A 2-parameter filtration using structuring elements $B_i$ defined in \eqref{eqn:1, 4, 9, 16,... rectangles}.   The notation $X^{\mathcal{O},\mathcal{C}}_{(-i,j)}$ denotes applying opening by $B_j$ followed by closing by $B_i$ on the original image level set $f^{-1}(0)$ depicted in (a) and formal definition can be found in Definition~\ref{def:multiparameter set}.  Note that the original image is the same as in Figure  \ref{fig:conceptual}(a) and the top row represents the one-filtration studied in Figure \ref{fig:conceptual}.}
	\label{fig:conceptualbiFiltration}	
\end{figure}

Figure \ref{fig:conceptualbiFiltration} shows sample images from this opening/closing bifiltration as applied to the Kanji example with additive noise also shown in Figure \ref{fig:conceptual}(a).  As seen in Figure \ref{fig:conceptual}, opening operations alone will not allow us to remove the small scale features due to additive noise while preserving the topology of the larger scale features.  By visual inspection of the bifiltration depicted in Figure \ref{fig:conceptualbiFiltration}, $X^{\mathcal{O},\mathcal{C}}_{(-2,3)}$ appears to be the most accurate rendering of the underlying Kanji image.  Since $X^{\mathcal{O},\mathcal{C}}_{(0,0)} \not\subseteq X^{\mathcal{O},\mathcal{C}}_{(-2,3)}$, there is no way to compare $X^{\mathcal{O},\mathcal{C}}_{(-2,3)}$ directly to the original image $X^{\mathcal{O},\mathcal{C}}_{(0,0)}$ using a one-filtration.  However, multiple one-filtrations within the $2$-filtration may be used to ``connect'' the two sets.  
  
We discuss an approach for using persistent homology information to search for optimal renderings within multifiltrations, along with extensions of multifiltrations from Section \ref{sec:ph} for binary images to a larger multifiltration that handles grayscale images, in Section \ref{sec:denoising algorithm}.

We now present a general multiparameter persistence framework using morphological operations. Consider a sequence of operations $\mathcal{E}_i:\mathcal{I}_P \rightarrow \mathcal{I}_P$ (e.g. erosion) and a sequence of operations, $\mathcal{D}_i:\mathcal{I}_P \rightarrow \mathcal{I}_P$ (e.g. dilation) satisfying the following: for any $f,~g\in\mathcal{I}_P$ and $i,j\in  \{0, 1,2,\dots,n\}$,

\begin{enumerate}
    \item[(A1)] if  $f \leq g$, then $\mathcal{E}_i(f) \leq \mathcal{E}_i(g)$ and $\mathcal{D}_i(f) \leq \mathcal{D}_i(g)$;  %For any $i$ and $f,g \in \mathcal{I}_P$ satisfy $f \leq g$, $\mathcal{E}_i(f) \leq \mathcal{E}_i(g)$ and $\mathcal{D}_i(f) \leq \mathcal{D}_i(g)$.
    \item[(A2)] if  $i \leq j$, then $\mathcal{E}_i(g) \geq \mathcal{E}_j(g)$ and $\mathcal{D}_i(g) \leq \mathcal{D}_j(g)$; %For any $i \leq j$ and $g \in \mathcal{I}_P$, $\mathcal{E}_i(g) \geq \mathcal{E}_j(g)$ and $\mathcal{D}_i(g) \leq \mathcal{D}_j(g)$.
    \item[(A3)] $\mathcal{D}_0(g) = \mathcal{E}_0(g) = g$.\vspace{0.2cm}
\end{enumerate}

When $\mathcal{E}_i= O_{B_i}$ and $\mathcal{D}_i= C_{B_i}$, assumptions (A1), (A2), and (A3) are similar to the \textit{sieving axioms} in granulometry: anti-extensivity, increasingness, and the absorption property (\cite{Soille2003,serra1984image}).  Assumption (A1) is the {\em increasing property} seen also in Proposition \ref{prop. dilation and erosion are increasing}.  Assumption (A2) is the {\em absorption property} (\cite{Najman-Mathematical-Morphology} Sec. 1.2.6, p.20).  Finally, combining assumptions (A2) and (A3) would lead to the anti-extensive or extensive property.

In what follows, let $\mathcal{E}_i$ and $\mathcal{D}_i$, $i\in \{0, 1,2,\dots,n\}$ be sequences satisfying (A1), (A2), and (A3). Consider $i\in\{0,\pm 1, ...,\pm n \}$ and define a function $M^{\mathcal{E}, \mathcal{D}}_i : \mathcal{I}_P \rightarrow \mathcal{I}_P$ as
\begin{equation}
\label{equ:op M}
M_i^{\mathcal{E}, \mathcal{D}}(g) := \left\{ \begin{array}{ll}
\mathcal{E}_i(g) & \mbox{for $i \geq 0$}\\
\mathcal{D}_{|i|}(g) & \mbox{for $i<0$}
\end{array}\right. ,
\end{equation}
and denote the $0$-level set by
\begin{equation}
\label{eqn:set rep-peter}
X_i^{\mathcal{E}, \mathcal{D}}(g) := \{ \bfx \in P \ | \ M_i^{\mathcal{E}, \mathcal{D}}(g)(\bfx)=0 \} = M_i^{\mathcal{E}, \mathcal{D}}(g)^{-1}(0).  
\end{equation}

When the context is understood, we sometimes abbreviate $M^{\mathcal{E}, \mathcal{D}}_i$ as $M_i$, and $X^{\mathcal{E}, \mathcal{D}}_i(g)$ as $X_i$.  The notation \eqref{equ:op M} unifies the operators $\mathcal{E}$ and $\mathcal{D}$ in the following way.

\begin{lemma}
\label{Lemma : Contravariant Order indeuced by M}
Let $i,j \in\{0,\pm 1, ...,\pm n \}$ and $g\in \mathcal{I}_P$ be a binary image.  Suppose $\mathcal{E}_i$ and $\mathcal{D}_i$ satisfy (A1), (A2), and (A3).
If $i \leq j$, then $M_j^{\mathcal{E}, \mathcal{D}}(g) \leq M_i^{\mathcal{E}, \mathcal{D}}(g)$.
\end{lemma}
\begin{proof}
Let $i\leq j$.  Suppose first that $i\geq 0$.  Then $M_j^{\mathcal{E}, \mathcal{D}}(g) = \mathcal{E}_j(g) \leq \mathcal{E}_i(g) = M_i^{\mathcal{E}, \mathcal{D}}(g)$.   If, on the other hand, $i < 0$, then there are two cases. In the case when $j > 0$, then $M_j^{\mathcal{E}, \mathcal{D}}(g) = \mathcal{E}_j(g) \leq g \leq \mathcal{D}_{|i|}(g) = M_i^{\mathcal{E}, \mathcal{D}}(g)$. In the second case where $j\leq 0$, then since $i \leq j \leq 0$, $|j| \leq |i|$ and we obtain $M_j^{\mathcal{E}, \mathcal{D}}(g) = \mathcal{D}_{|j|}(g) \leq \mathcal{D}_{|i|}(g) = M_i^{\mathcal{E}, \mathcal{D}}(g)$. 
\end{proof}

The essential step in obtaining a multi-parameter filtration is to apply $M^{\mathcal{E},\mathcal{D}}$ inductively.  This requires us to  extend the notation of \eqref{equ:op M} and \eqref{eqn:set rep-peter} to a multi-index $\mathbf{i}\in\mathbb{Z}^k$.

\begin{definition}
\label{def:multiparameter set}
Let $\{\mathcal{E}_i\}$ and $\{\mathcal{D}_i\}$ be sequences of morphological operations that satisfy (A1), (A2), and (A3). For $k, n \in \mathbb{N}$ and $\mathbf{i} = (i_1, i_2, \dots, i_k) \in \{0, \pm 1,..., \pm  n \}^k$, we define $M^{\mathcal{E},\mathcal{D}}_{\mathbf{i}}:\mathcal{I}_P \rightarrow \mathcal{I}_P$ and $X^{\mathcal{E},\mathcal{D}}_{\mathbf{i}}\subseteq P$ by
\begin{equation}
    M^{\mathcal{E},\mathcal{D}}_{\mathbf{i}}(g) = (M^{\mathcal{E},\mathcal{D}}_{i_1} \circ M^{\mathcal{E},\mathcal{D}}_{i_2} \circ \cdots \circ M^{\mathcal{E},\mathcal{D}}_{i_k})(g),
\end{equation}
and
\begin{equation}
\label{eqn:set rep gen-peter}
    X^{\mathcal{E},\mathcal{D}}_{\mathbf{i}}(g) = \{ \bfx \in P \ | \ M^{\mathcal{E},\mathcal{D}}_{\mathbf{i}}(g)(\bfx) = 0 \} = M^{\mathcal{E},\mathcal{D}}_{\mathbf{i}}(g)^{-1}(0).
\end{equation}
Similarly, we abbreviate the notation $M^{\mathcal{E},\mathcal{D}}_{\mathbf{i}}$ as $M_{\mathbf{i}}$ and $X^{\mathcal{E},\mathcal{D}}_{\mathbf{i}}(g)$ as $X_{\mathbf{i}}$ if operations $\mathcal{E}, \mathcal{D}$ and image $g$ are specified.
\end{definition}

For example, for ${\mathbf{i}}=(-1,1)$, $M_{\mathbf{i}}(g) = M_{-1}(M_1(g))$ means that the image $g$ is filtered by $\mathcal{E}_1$ followed by $\mathcal{D}_1$, i.e. $M_{\mathbf{i}}(g) = M_{-1} \circ M_1(g) = \mathcal{E}_1 (\mathcal{D}_1(g) )$.

Motivated by \eqref{equ:bifiltration O and C}, we consider the sets $X_{\mathbf{i}}$ formed by the application of alternating $\mathcal{E}$ and $\mathcal{D}$ operations.  Using \eqref{equ:op M}, we see that alternating these operations corresponds to a multi-index consisting of an alternating sequence of integers.

\begin{definition}\label{def:alt_seq}
The sequence $\mathbf{i}=(i_1, i_2, \ldots, i_k)\in \{0, \pm 1,..., \pm  n \}^k$ is an \textit{alternating sequence} if $i_{l} \cdot i_{l+1} \leq 0$ for all $l \in \{ 1,2, ..., k\}$. By extension, the set $A\subseteq  \{0, \pm 1,..., \pm  n \}^k$ is a \textit{set of alternating sequences} if it contains only \textit{alternating sequences}.
\end{definition}

We are now ready to present our main theorem: alternating the operations $\mathcal{E}$ and $\mathcal{D}$ leads to a multi-parameter filtration.
\begin{theorem}
\label{thm:multifiltration-peter}
%Let $m \in \bbN$, $P \subseteq \mathbb{Z}^m$, $g \in \mathcal{I}_P$, and 
Let $g$ be a binary image, and $A \subseteq \{0,\pm 1, ...,\pm n \}^k$ be a set of alternating sequences. Assume  $\mathcal{E}_i(g)$, $\mathcal{D}_i(g) : \mathcal{I}_P \rightarrow \mathcal{I}_P$, $i\in \{1,2,\dots,n\}$ satisfy (A1), (A2) and (A3). Then $\left\{X^{\mathcal{E}, \mathcal{D}}_{\mathbf{i}}\right\}_{\mathbf{i}\in A}$ is a $k$-parameter filtration. 
\end{theorem}
\begin{proof}
Let $\textbf{u} = (u_1, ..., u_n), \textbf{v} = (v_1, ..., v_n) \in A$ and $\textbf{u} \leq \textbf{v}$.  By Definition \ref{def:multi filtration}, we need to verify that $M_{\textbf{u}}(g)^{-1}(0) \subseteq M_{\textbf{v}}(g)^{-1}(0)$. 

By Lemma \ref{Lemma : Contravariant Order indeuced by M}, since $u_n \leq v_n$ we have that $M_{v_n}(g) \leq M_{u_n}(g)$. Applying (A1), we see that
\begin{equation}
\label{equ:proof main thm step1}
    (M_{v_{n-1}} \circ M_{v_n})(g) \leq (M_{v_{n-1}} \circ M_{u_n})(g).
\end{equation}
Since $u_{n-1} \leq v_{n-1}$ by Lemma \ref{Lemma : Contravariant Order indeuced by M} again, we have 
\begin{equation}
\label{equ:proof main thm step2}
    (M_{v_{n-1}} \circ M_{u_n})(g) \leq (M_{u_{n-1}} \circ M_{u_n})(g).
\end{equation}
Therefore, by combining \eqref{equ:proof main thm step1} and \eqref{equ:proof main thm step2}, we prove that $ (M_{v_{n-1}} \circ M_{v_n})(g) \leq (M_{u_{n-1}} \circ M_{u_n})(g)$.
Finally, by applying the argument inductively one may conclude that
\begin{equation*}
\begin{split}
    M_{\textbf{v}}(g) &= (M_{v_1} \circ \cdots \circ M_{v_n})(g) \leq (M_{u_1} \circ \cdots \circ M_{u_n})(g) = M_{\textbf{u}}(g).
\end{split}
\end{equation*}
By Proposition \ref{Prop. Equivalent property of g <= f}, we conclude that $M_{\textbf{u}}(g)^{-1}(0) \subseteq M_{\textbf{v}}(g)^{-1}(0)$.
\end{proof}

\begin{remark}
\label{Remark : Why alternating seq?}
For purposes of exposition and to align with common practices in using morphological operations in image smoothing, we focused the composition of operations on alternating sequences (see \cite{serra1984image,Najman-Mathematical-Morphology,Soille2003}).  This is inherent in (A2) as well as the stipulation that the indexing set $A$ in Theorem \ref{thm:multifiltration-peter} consists of alternating sequences.  We note here, however, that the theorem holds true even if $A$ contains sequences that are not alternating.
\end{remark}

We now discuss examples to illustrate the framework given in Theorem \ref{thm:multifiltration-peter}. As a first example, consider erosion and dilation given as $\mathcal{E}_i := \epsilon_{B_i}$ and $\mathcal{D}_i := \delta_{B_i}$.  For this pair of operations, (A1) follows from Proposition \ref{prop. dilation and erosion are increasing}, (A2) follows from Proposition \ref{Prop:erosion dilation subset}, and (A3) is clear. 

Therefore, by Theorem~\ref{thm:multifiltration-peter}, $\left\{X^{\epsilon, \delta}_{\mathbf{i}}\right\}_{\mathbf{i}\in A}$ forms a multi-parameter filtration, where $A$ is any set of alternating sequences. 
As a second example, consider the opening and closing operations and let  $\mathcal{E}_i := O_{B_i}$ and $\mathcal{D}_i := C_{B_i}$. By Theorem~\ref{thm:multifiltration-peter}, $\left\{X^{\mathcal{O},\mathcal{C}}_{\mathbf{i}}\right\}_{\mathbf{i}\in A}$ forms a multi-parameter filtration. In fact, erosion/closing, and opening/dilation would also lead to multiparameter filtrations. It is important to note that while erosion/dilation and opening/closing lead naturally to multiparameter filtrations, the top-hat transformations do not. As mentioned in Section \ref{sec:ph}, these transformations do not satisfy (A1) and (A3) in general and, therefore, do not satisfy the hypotheses of Theorem \ref{thm:multifiltration-peter}.

At this point, $g$ is assumed to be a binary image.  If $g$ is a grayscale image, one may combine the sublevel set filtration with the multiparameter filtration described in Theorem \ref{thm:multifiltration-peter} to obtain another multiparameter filtration. In the rest of this section, we will formulate this concept as an extension of Theorem~\ref{thm:multifiltration-peter}.

Let $\{ \mathcal{E}_i \}_{i = 1}^n$ and $\{ \mathcal{D}_i \}_{i = 1}^n$ be sequences of operations $\mathcal{I}_P \rightarrow \mathcal{I}_P$ satisfying $(A1)$, $(A2)$ and $(A3)$. Moreover, we also require that for all $i \in \{ 1,2, ..., n\}$ and $t \in \{ 0,1,,2 ..., N \}$,
\begin{enumerate}
    \item[(A4)]$\mathcal{E}_i \circ \tau_t = \tau_t \circ \mathcal{E}_i$ and $\mathcal{D}_i \circ \tau_t = \tau_t \circ \mathcal{D}_i$.
\end{enumerate}
This assumption means that the morphological operations and thresholding operation commute. Proposition \ref{Proposition : Ristriction and opening/closing are commutative} shows that $\delta$, $\epsilon$, $O$, $C$ satisfy (A4).

For every $\bfu \in \{ 0, \pm 1, ..., \pm n\}^k$ let
$X_{t,\mathbf{u}}^{\mathcal{E},\mathcal{D}} := M_{\mathbf{u}}(f)_t^{-1}(0).$  We now show that if (A1)-(A4) are satisfied, then $\{X_{t,\mathbf{u}}^{\mathcal{E},\mathcal{D}}\}_{(t, \mathbf{u})}$ forms a $(k+1)$-parameter filtration.  To achieve that, we need to verify that $M_{\mathbf{v}}(f_s) \leq M_{\mathbf{u}}(f_t)$, for all $(t,\mathbf{u}) \leq (s, \mathbf{v})$.
By (A4), we have $M_{\bfu}(f)_t = M_{\bfu}(f_t)$ for $t \in \{ 0,1,,2 ..., N \}$. Therefore, by (A4) and Theorem  \ref{thm:multifiltration-peter}, one has
\begin{equation*}
    M_{\mathbf{v}}(f_s) \leq M_{\mathbf{u}}(f_s) =  M_{\mathbf{u}}(f)_s \leq  M_{\mathbf{u}}(f)_t = M_{\mathbf{u}}(f_t).
\end{equation*}
%for all thresholds $t$ by Theorem \ref{thm:multifiltration-peter}. 
We summarize the above discussion into the following result.
\begin{corollary} 
\label{thm:multifiltration2-peter}
Let $g$ be a grayscale image, and $A \subseteq \{0,\pm 1, ...,\pm n \}^k$ be a set of alternating sequences.  Assume  $\mathcal{E}_i(g)$, $\mathcal{D}_i(g) : \mathcal{I}_P \rightarrow \mathcal{I}_P$, $i\in \{1,2,\dots,n\}$ satisfy (A1), (A2), (A3), and (A4). Then $\{X^{\mathcal{E},\mathcal{D}}_{t,\mathbf{u}}\}_{(t,\mathbf{u})}$ is a $(k+1)$-parameter filtration.
\end{corollary}

The following is an example of the framework in Theorem \ref{thm:multifiltration2-peter},
\begin{equation}
	\label{equ:bifiltration}
	\begin{tabular}{ccccccc}
		$X_{1,(n, -2, 3)}^{\mathcal{E},\mathcal{D}}$ & $\subseteq$ & $X_{2,(n, -2, 3)}^{\mathcal{E},\mathcal{D}}$ & $\subseteq\cdots\subseteq$ & $X_{T-1,(n, -2, 3)}^{\mathcal{E},\mathcal{D}}$ & $\subseteq$ & $X_{T,(n, -2, 3)}^{\mathcal{E},\mathcal{D}}$  \\
		\rotatebox[origin=c]{90}{$\subseteq$} &  & \rotatebox[origin=c]{90}{$\subseteq$} & $\vdots$& \rotatebox[origin=c]{90}{$\subseteq$} &  & \rotatebox[origin=c]{90}{$\subseteq$} \\	
		$X_{1,(n-1, -2, 3)}^{\mathcal{E},\mathcal{D}}$ & $\subseteq$ & $X_{2,(n-1, -2, 3)}^{\mathcal{E},\mathcal{D}}$ & $\subseteq\cdots\subseteq$ & $X_{T-1,(n-1, -2, 3)}^{\mathcal{E},\mathcal{D}}$ & $\subseteq$ & $X_{T,(n-1
		, -2, 3)}^{\mathcal{E},\mathcal{D}}$  \\
		\rotatebox[origin=c]{90}{$\subseteq$} &  & \rotatebox[origin=c]{90}{$\subseteq$} & $\vdots$& \rotatebox[origin=c]{90}{$\subseteq$} &  & \rotatebox[origin=c]{90}{$\subseteq$} \\			
		$\vdots$ &  & $\vdots$ & $\vdots$& $\vdots$&& $\vdots$ \\			
		\rotatebox[origin=c]{90}{$\subseteq$} &  & \rotatebox[origin=c]{90}{$\subseteq$} & $\vdots$& \rotatebox[origin=c]{90}{$\subseteq$} &  & \rotatebox[origin=c]{90}{$\subseteq$} \\	
		$X_{1,(1, -2, 3)}^{\mathcal{E},\mathcal{D}}$ & $\subseteq$ & $X_{2,(1, -2, 3)}^{\mathcal{E},\mathcal{D}}$ & $\subseteq\cdots\subseteq$ & $X_{T-1,(1, -2, 3)}^{\mathcal{E},\mathcal{D}}$ & $\subseteq$ & $X_{T,(1, -2, 3)}^{\mathcal{E},\mathcal{D}}$  \\
		\rotatebox[origin=c]{90}{$\subseteq$} &  & \rotatebox[origin=c]{90}{$\subseteq$} & $\vdots$& \rotatebox[origin=c]{90}{$\subseteq$} &  & \rotatebox[origin=c]{90}{$\subseteq$} \\	
		$X_{1,(0, -2, 3)}^{\mathcal{E},\mathcal{D}}$ & $\subseteq$ & $X_{2,(0, -2, 3)}^{\mathcal{E},\mathcal{D}}$ & $\subseteq\cdots\subseteq$ & $X_{T-1,(0, -2, 3)}^{\mathcal{E},\mathcal{D}}$ & $\subseteq$ & $X_{T,(0, -2, 3)}^{\mathcal{E},\mathcal{D}}$ 
	\end{tabular}.
\end{equation}

While different methods, including the rank invariant function \cite{carlsson2009theory} and sheaf theory \cite{KashiwaraSchapira2018,10.1093/imrn/rnz145}, have been developed to study multi-parameter persistence, for purposes of illustration we will focus on computing persistent homology along {\em nondecreasing paths} in the constructed multi-filtration.

\begin{definition}\label{def:nondecreasing}  Define a {\em nondecreasing path} in indexing set $A$ as a sequence $\mathbf{u}_0, \mathbf{u}_1, \ldots, \mathbf{u}_l \in A$ such that $\mathbf{u}_i \leq \mathbf{u}_{i+1}$ for all $i=0,\ldots, l$.  Then for a multifiltration $\{X_{\mathbf{u}}\}_{\mathbf{u}\in A}$ and nondecreasing path $\mathbf{u}_0, \mathbf{u}_1, \ldots, \mathbf{u}_l$ in $A$, $\{X_{\mathbf{u}_i}\}_{i}$ is a one-parameter filtration.
\end{definition}
As we outline in the following section, this structure allows us to systematically extract information about geometric scale and optimize for certain topological features.  Following multiple or successive nondecreasing paths allows for greater exploration of the multifiltration.

% %%%%%%%%%%%%%%%%%%%%%%%%%%%%%%%%%%%%%%%%%%%%%%%%%%%%%%%%%%%%%%%%%%%%%%%%%%%%%%%%%%%%%%%%%
% %%%%%%%%%%%%%%%%%%%%%%%%%%%%%%%%%%%%%%%%%%%%%%%%%%%%%%%%%%%%%%%%%%%%%%%%%%%%%%%%%%%%%%%%%
% %%%%%%%%%%%%%%%%%%%%%%%%%%%%%%%%%%%%%%%%%%%%%%%%%%%%%%%%%%%%%%%%%%%%%%%%%%%%%%%%%%%%%%%%%
% %%%%%%%%%%%%%%%%%%%%%%%%%%%%%%%%%%%%%%%%%%%%%%%%%%%%%%%%%%%%%%%%%%%%%%%%%%%%%%%%%%%%%%%%%

%%%%%%%%%%%%%%%%%%%%%%%%%%%%%%%%%%%%%%%%%%%%%%%%%%%%%%%%%%%%%%%%%%%%%%%%%%%%%%%%%%%%

%%%%%%%%%%%%%%%%%%%%%%%%%%%%%%%%%%%%%%%%%%%%%%%%%%%%%%%%%%%%%%%%%%%%%%%%%%%%%%%%%%%%
\section{Application: A Denoising Algorithm for Salt and Pepper noise}
\label{sec:denoising algorithm}
We now use the multiparameter filtration to construct a denoising algorithm aimed at removing salt and pepper (small spatial scale, high amplitude) noise.  See, e. g., Figure~\ref{fig:conceptual} where one goal is removing small scale white regions from the images in order to focus on the larger scale features.
%The algorithm is motivated by the firn application in Section \ref{sec:firn} where one goal is removing ice dust resulting from sample collection from the images in order to focus on the larger scale features. 
Given a binary image $f$, we wish to apply alternating opening/closing operations to $f$.  Traditionally, this requires visual inspection to tune the size of the utilized structural elements as well as the number of operations performed.  %This is indeed the approach in Section \ref{sec:firn}. 
We now seek to automate this process by more fully utilizing the full multiparameter persistence framework. In this section, we will describe details of our proposed algorithm, demonstrate it on synthetic images, and extend it and apply it to grayscale and color images.

\begin{figure}
    \centering
    \subfloat[Ground truth.]{
    \includegraphics[width=0.25\linewidth]{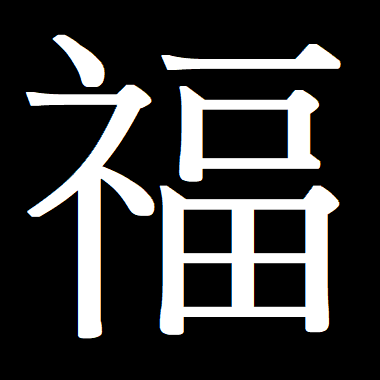}} \\
    \subfloat[Noise in density $0.4$.]{
    \includegraphics[width=0.25\linewidth]{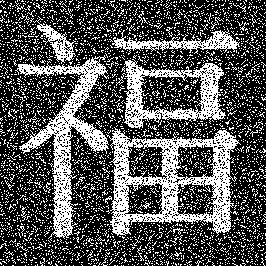}} \quad
    \subfloat[Algorithm \ref{Alg:denoise} in density $0.4$.]{%\subfloat[Denoised image in density $0.4$.]{
    \includegraphics[width=0.25\linewidth]{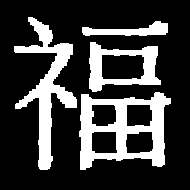}} \quad
    \subfloat[{\tt{denoiseImage}} in density $0.4$.]{
    \includegraphics[width=0.25\linewidth]{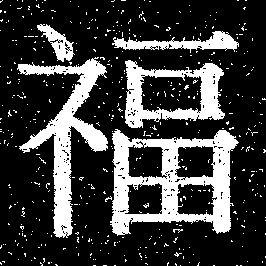}}\\    
    \subfloat[Noise in density $0.7$.]{
    \includegraphics[width=0.25\linewidth]{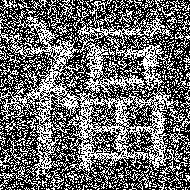}} 
 \quad
    \subfloat[Algorithm \ref{Alg:denoise} in density $0.7$.]{%\subfloat[Denoised image in density $0.7$.]{
    \includegraphics[width=0.25\linewidth]{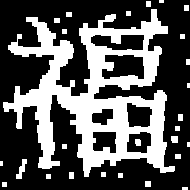}}
\quad
    \subfloat[{\tt{denoiseImage}} in density $0.7$.]{
    \includegraphics[width=0.25\linewidth]{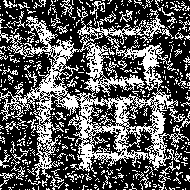}}\\
    \caption{Conceptual images with salt and pepper noise and the results of different denoised algorithms: Algorithm \ref{Alg:denoise}, and {\tt{denoiseImage}}.  The parameters for Algorithm \ref{Alg:denoise} are {\tt{MaxIter}=10} and {\tt {Sizetol}=5}. The resulting alternating opening/closing sequences $\bfu$ of (c) and (f) are $(-2,4,-1,3,-1,2,-1,1,-1)$ and $(-4,4,-1,3,-1,2,-1,1,-1)$ respectively.
    } 
    \label{fig:Character}
\end{figure}

The proposed algorithm is iterative.  In each iteration, we will use persistence diagrams computed along a nondecreasing path in the multiparameter filtration to guide the choice of a structuring element used for opening or closing.  To get started, consider a binary image $f$ contaminated by salt and pepper noise, {that is, certain pixels have been switched to either white (salt) or  black (pepper).  See Figure~\ref{fig:Character} (b) and (e) for examples of images contaminated by the salt and pepper noise.}  Each iteration consists two steps: closing operation followed by opening.  It is also possible to perform opening operation followed by closing for instead.  %In what follows, we will perform closing followed by opening.

We first consider the closing filtration of $f$, $\{X^{\mathcal{C}}_j\}_{j=-n}^0$, as shown in \eqref{equ:closing filtration}, where in this case $X^{\mathcal{C}}_j = C_{B_{|j|}}(f)^{-1}(0)$.  By construction, $X^{\mathcal{C}}_0 = f^{-1}(0)$ is the largest set in the closing filtration and the persistence diagram can be decomposed into
\begin{equation*}
    \mathcal{P}_0(\{X^{\mathcal{C}}_j\}_{j=-n}^0) = \mathcal{P}_0^{\mathcal{C}} \cup  \{ (b,d)\in \mathcal{P}_0|~ d<0 \},
\end{equation*}
where $\mathcal{P}_0^{\mathcal{C}} := \{ (b,d)\in \mathcal{P}_0|~ d=0 \}$.  Similar to the discussion for opening filtration in Section \ref{sec:ph}, $\mathcal{P}_0^{\mathcal{C}}$ contains features (black regions) that are present in the original image $X^{\mathcal{C}}_0=f^{-1}(0)$, and $|b|$ where $b\in \mathcal{P}_0^{\mathcal{C}}$ indicates the size of the feature by giving the amount of closing required to remove it from the image.  Since salt and pepper noise creates features that are small in spatial scale, we take a conservative route by choosing 
\begin{equation*}
    i_c =  \min \{ |b| ~|~ (b,0) \in \mathcal{P}_0^{\mathcal{C}} \}.
\end{equation*}
The binary image of the first step is then $X_{i_c}:=C_{B_{i_c}}(f)^{-1}(0)$.  To generalize this approach, we note that a gap in the death coordinate values in $\mathcal{P}_0^{\mathcal{C}}$ can be used to detect a separation in spatial scales for features in the original image.

Using the new binary image $C_{B_{i_c}}(f)$, we now consider the opening filtration of $C_{B_{i_c}}(f)$, $\{X^{\mathcal{O}}_i\}_{i=0}^n$, as shown in \eqref{equ:opening filtration}, where in this case $X^{\mathcal{O}}_i = O_{B_i}(C_{B_{i_c}}(f))^{-1}(0)$.   As discussed in Section \ref{sec:ph}, $\mathcal{P}_1(\{X^{\mathcal{O}}_i\}_{i=0}^n)$ reveals size information of the white regions.  Specifically, $\mathcal{P}_1(\{X^{\mathcal{O}}_i\}_i)$ can be decomposed as
\begin{equation*}
    \mathcal{P}_1(\{X^{\mathcal{O}}_i\}_i) = \mathcal{P}_1^{\mathcal{O}} \cup  \{ (b,d)\in \mathcal{P}_1|~ b>0 \},
\end{equation*}
where $\mathcal{P}_1^{\mathcal{O}}:= \{ (b,d)\in \mathcal{P}_1|~ b=0 \}$.  As demonstrated in Section \ref{sec:ph}, $\mathcal{P}_1^{\mathcal{O}}$ contains features that are present in the original binary image, $X^{\mathcal{O}}_0$, and $d\in \mathcal{P}_1^{\mathcal{O}}$ indicates the spatial size of the feature, that is, the amount of opening required to remove the feature.   Similar to our approach in the first step using opening, we choose the size of the structuring element for closing to be
\begin{equation*}
    i_o = \min \{d~|~(0,d) \in \mathcal{P}_1^{\mathcal{O}} \}.
\end{equation*}
The binary image following the second step is now $X_{(i_o, i_c)} = O_{B_{i_o}}(C_{B_{i_c}}(f))^{-1}(0)$.

\begin{figure}
\begin{equation*}
    \xymatrix@-1.4em{
	&& X \ar[rr]\ar[rr]\ar[rr]\ar[rr]^{C_{B_1}}\ar[rr]\ar[rr]\ar[rr]\ar[rr]\ar@{--}@[red][dd]_(.5){\color{red} \rotatebox[origin=c]{90}{$\supseteq$}}
	&& X_{(-1)} \ar[dd]^(.55){O_{B_1}}\ar[dd]\ar[dd]\ar[dd]\ar[dd]\ar[dd] \ar@{--}@[red][rr]^(.5){\color{red} \rotatebox[origin=c]{360}{$\supseteq$}} 
	&& X_{(-2)} \ar@{--}@[red][dd]^(.5){}\\
	&&& X_{(1,-1)} \ar@{--}@[blue][dd]_(.5){\color{blue} \rotatebox[origin=c]{90}{$\supseteq$}}
	\ar@{--}@[blue][rr]_(.7){\color{blue} \rotatebox[origin=c]{360}{$\supseteq$}}
	\ar@/^1pc/[rrrr]^{C_{B_2}}{ }
	\ar@/^1pc/[rrrr]^{ }{ }
	\ar@/^1pc/[rrrr]^{ }{ }
	\ar@/^1pc/[rrrr]^{ }{ }
	&& X_{(-1,1,-1)} \ar@{--}@[blue][rr]_(.6){\color{blue} \rotatebox[origin=c]{360}{$\supseteq$}}
	\ar@{--}@[blue][dd] 
	&& X_{(-2,1,-1)} \ar@/^2.5pc/[dddd]^(.5){O_{B_2}} \ar@/^2.5pc/[dddd] \ar@/^2.5pc/[dddd] \ar@/^2.5pc/[dddd] \ar@/^2.5pc/[dddd] \ar@/^2.5pc/[dddd] \ar@{--}@[blue][dd]^(.5){} \\
	&& X_{(1)} \ar@{--}@[red][dd]_(.5){\color{red} \rotatebox[origin=c]{90}{$\supseteq$}}\ar@{--}@[red][rr]|{ }
	&& X_{(1,-1)} 
	\ar[ul]_(.45){=}\ar[ul]_(.45){=}\ar[ul]\ar[ul]\ar[ul]_(.45){=} \ar@{--}@[red][dd]^(.5){} \ar@{--}@[red][rr]^(.5){} 
	&& X_{(1,-2)} \ar@{--}@[red][dd]^(.5){}\\
	&&& X_{(1,1,-1)} \ar@{--}@[blue][dd]_(.5){\color{blue} \rotatebox[origin=c]{90}{$\supseteq$}} \ar@{--}@[blue][rr]|\hole^(.36){ } 
	&& X_{(1,-1,1,-1)} \ar@{--}@[blue][rr] \ar@{--}@[blue][dd]^(.5){} 
	&& X_{(1,-2,1,-1)} \ar@{--}@[blue][dd]^(.5){}\\
	&& X_{(2)} \ar@{--}@[red][rr]|{ }
	&& X_{(2,-1)} \ar@{--}@[red][rr] 
	&& X_{(2,-2)}\\
	&&& X_{(2,1,-1)}\ar@{--}@[blue][rr] 
	&& X_{(2,-1,1,-1)} \ar@{--}@[blue][rr] && X_{(2,-2,1,-1)}\\
}
\end{equation*}
%\centering
    % \subfloat[$X$]{
    % \includegraphics[width=0.28\linewidth]{figures/kanji.png}}
    % \subfloat[$X_{(-2,2,-1,1)}$]{
    % \includegraphics[width=0.28\linewidth]{ICCV2019/figures/kanji_After.png}}
    \caption{Cartoon illustration of steps in Algorithm \ref{Alg:denoise} in the multiparameter filtration that would produce the alternating sequence $(2, -2, 1, -1)$. Red dotted lines highlight a bifiltration layer, and blue dotted lines highlight a different bifiltration layer.  The black solid line represents the path and selections made by Algorithm \ref{Alg:denoise}. %In this demonstration, $B_1=S(3)$ and $B_2=S(4)$ are $3 \times 3$ and $4 \times 4$ squares defined in~\eqref{eqn:1, 4, 9, 16,... rectangles}. {\color{red}Is the notation $B_i$ consistent?}
    }
    \label{fig:scheme of algorithm}
\end{figure}
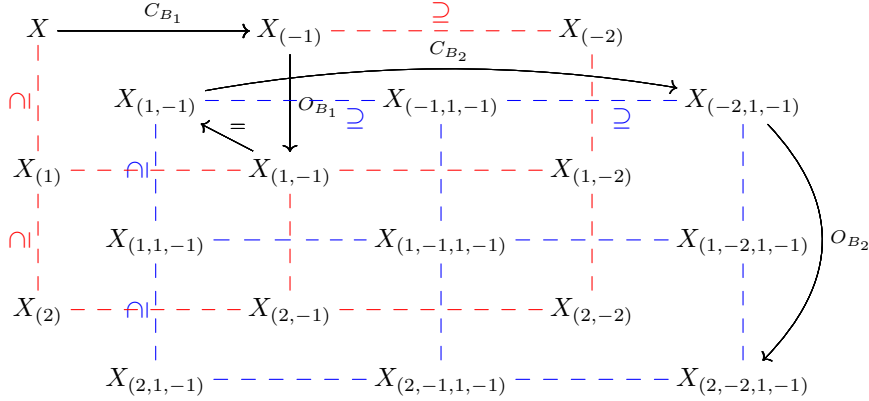

We repeat this alternating process. The stopping criterion is when the selected structuring element size exceeds a preset maximum, ${\tt SizeTol}$.  This could be given by the size of the image, or given as an upper bound on the spatial size of noisy features or features we wish to remove.  %For firn, this is an upper bound on the largest size of a piece of the ice dust due to core extraction, as specified by a domain expert.  
The algorithm is summarized in Algorithm \ref{Alg:denoise}. Figure \ref{fig:scheme of algorithm} illustrates Algorithm \ref{Alg:denoise} in a schematic way in the multiparameter space.

%%%%%%%%%%%%%%%%%%%%%%%%%%%%%%%%%%%%%%%%%%%%%%%%%%%%%%%%%%%%%%%%%%%%%%%%%%%%%%%%%%%%%%%%%%%%%%%%%%%%%%%%%%%%%%%%%%%%%%%%%%%%%
%%%%%%%%%%%%%%%%%%%%%%%%%%%%%%%%%%%%%%%%%%%%%%%%%%%%%%%%%%%%%%%%%%%%%%%%%%%%%%%%%%%%%%%%%%%%%%%%%%%%%%%%%%%%%%%%%%%%%%%%%%%%%
%%%%%%%%%%%%%%%%%%%%%%%%%%%%%%%%%%%%%%%%%%%%%%%%%%%%%%%%%%%%%%%%%%%%%%%%%%%%%%%%%%%%%%%%%%%%%%%%%%%%%%%%%%%%%%%%%%%%%%%%%%%%%
%%%%%%%%%%%%%%%%%%%%%%%%%%%%%%%%%%%%%%%%%%%%%%%%%%%%%%%%%%%%%%%%%%%%%%%%%%%%%%%%%%%%%%%%%%%%%%%%%%%%%%%%%%%%%%%%%%%%%%%%%%%%%

\begin{algorithm}[H]
\SetAlgoLined
 {\tt{Input}:} A binary image $f \in \mathcal{I}_P$ and its set companion $X=f^{-1}(0)$, $\{ \mathbf{0} \} \subseteq_S B_1 \subseteq_S \cdots \subseteq_S B_n$, stopping parameter {\tt SizeTol}, and maximum number of iterations \tt{MaxIter}.
%\KwInput{A binary image $f \in \mathcal{I}_P$ and its set companion $X=f^{-1}(0)$, $\{ \mathbf{0} \} \subseteq_S B_1 \subseteq_S \cdots \subseteq_S B_n$, stopping parameter {\tt SizeTol}, and maximum number of iterations {\tt{MaxIter}.}

\textbf{Output:} $X^{\mathcal{O},\mathcal{C}}_{\mathbf{i}} =: X_{\mathbf{i}}$, where $\mathbf{i} \in \{ 0, \pm 1, \dots, \pm n \}^l$ for some $l \in \bbN$. 
%\KwOutput{$X^{\mathcal{O},\mathcal{C}}_{\mathbf{i}} =: X_{\mathbf{i}}$, where $\mathbf{i} \in \{ 0, \pm 1, \dots, \pm n \}^l$ for some $l \in \bbN$}
%}

Denote $X = f^{-1}(0)$. 

Set $g \leftarrow f$

\For{$j = 1,2,...,$ {\tt{MaxIter}} }{
Compute $\mathcal{P}_0(\{X^{\mathcal{C}}_i\}_{i=-n}^0)$

Find $i_c = \min \{|b|~ | (b,0) \in \mathcal{P}_0^{\mathcal{C}} \}$ 

\eIf{$i_c \leq {\tt SizeTol}$}{
   $X \leftarrow C_{B_{i_c+1}}(f)^{-1}(0)$ %\;
   $\mathbf{i} \leftarrow [i_c \quad \mathbf{i}]$
   }
   {Return $X$ and $\mathbf{i}$}

Compute $\mathcal{P}_1(\{X^{\mathcal{O}}_i\}_{i=0}^n)$

Find $i_o = \min \{d~ | (0,d) \in \mathcal{P}_1^{\mathcal{O}} \}$ 

\eIf{$i_o \leq {\tt SizeTol}$}{
   $X \leftarrow O_{B_{i_o}}(f)^{-1}(0)$ %\;
   $\mathbf{i} \leftarrow [i_o \quad \mathbf{i}]$
   }
   {Return $X$ and $\mathbf{i}$}
}
Return $X$ and $\mathbf{i}$
end

\caption{The proposed denoising algorithm .}
\label{Alg:denoise}
\end{algorithm}

%%%%%%%%%%%%%%%%%%%%%%%%%%%%%%%%%%%%%%%%%%%%%%%%%%%%%%%%%%%%%%%%%%%%%%%%%%%%%%%%%%%%%%%%%%%%%%%%%%%%%%%%%%%%%%%%%%%%%%%%%%%%%
%%%%%%%%%%%%%%%%%%%%%%%%%%%%%%%%%%%%%%%%%%%%%%%%%%%%%%%%%%%%%%%%%%%%%%%%%%%%%%%%%%%%%%%%%%%%%%%%%%%%%%%%%%%%%%%%%%%%%%%%%%%%%
%%%%%%%%%%%%%%%%%%%%%%%%%%%%%%%%%%%%%%%%%%%%%%%%%%%%%%%%%%%%%%%%%%%%%%%%%%%%%%%%%%%%%%%%%%%%%%%%%%%%%%%%%%%%%%%%%%%%%%%%%%%%%
%%%%%%%%%%%%%%%%%%%%%%%%%%%%%%%%%%%%%%%%%%%%%%%%%%%%%%%%%%%%%%%%%%%%%%%%%%%%%%%%%%%%%%%%%%%%%%%%%%%%%%%%%%%%%%%%%%%%%%%%%%%%%
%%%%%%%%%%%%%%%%%%%%%%%%%%%%%%%%%%%%%%%%%%%%%%%%%%%%%%%%%%%%%%%%%%%%%%%%%%%%%%%%%%%%%%%%%%%%%%%%%%%%%%%%%%%%%%%%%%%%%%%%%%%%%
%%%%%%%%%%%%%%%%%%%%%%%%%%%%%%%%%%%%%%%%%%%%%%%%%%%%%%%%%%%%%%%%%%%%%%%%%%%%%%%%%%%%%%%%%%%%%%%%%%%%%%%%%%%%%%%%%%%%%%%%%%%%%
%%%%%%%%%%%%%%%%%%%%%%%%%%%%%%%%%%%%%%%%%%%%%%%%%%%%%%%%%%%%%%%%%%%%%%%%%%%%%%%%%%%%%%%%%%%%%%%%%%%%%%%%%%%%%%%%%%%%%%%%%%%%%
%%%%%%%%%%%%%%%%%%%%%%%%%%%%%%%%%%%%%%%%%%%%%%%%%%%%%%%%%%%%%%%%%%%%%%%%%%%%%%%%%%%%%%%%%%%%%%%%%%%%%%%%%%%%%%%%%%%%%%%%%%%%%

%%%%%Numerical experiments
To test the proposed algorithm, we again use the $190 \times 190$ binary image shown in Figure \ref{fig:Character}(a) as the ground truth.  We add salt and pepper noise to the ground truth with various levels of noise densities.  The noise density parameter gives the portion, in probability, of pixels whose values have been changed from their original values (to either white or black).  We use the Matlab built-in function {\tt{imnoise}} along with a specified noise density parameter.  For instance, noisy images with noise density 0.4 and 0.7 can be found in Figure \ref{fig:Character}(b) and Figure \ref{fig:Character}(d), respectively.  The denoised images by Algorithm~\ref{Alg:denoise} can be found in Figure \ref{fig:Character}(c) and Figure \ref{fig:Character}(f), respectively;  the denoised images by {\tt{imnoise}} can be found in Figure \ref{fig:Character}(d) and Figure \ref{fig:Character}(g), respectively  Visually, the denoised images by Algorithm~\ref{Alg:denoise} are close to the ground truth.  Even in the case when noise density is 0.7, the denoised image (Figure \ref{fig:Character}(f)) still recovers much of the core structure of the ground truth Figure \ref{fig:Character}(a).  On the other hand, denoised images by {\tt{imnoise}} are still pixelated.

We conduct an experiment to further test the proposed algorithm.  Again, we take the ground truth image Figure~\ref{fig:Character}(a) and add salt and pepper noise to it with noise densities from $0.1$ to $1.0$.  For each noise density, we construct $1000$ noisy images with the prescribed noise density, and for each noisy image, we apply Algorithm~\ref{Alg:denoise} with {\tt{MaxIter}=10} and {\tt {Sizetol}=5}.  The metric we use to compare the ground truth with the computed one is the intersection over union (IOU) defined as $\frac{| S_1 \cap S_2 | }{| S_1 \cup S_2 |}$ for sets $S_1$ and $S_2$.  In this case, we use $S_1=f^{-1}(0)$ and $S_2=\widehat{f}^{-1}(0)$, the black sets for the ground truth image and the output, denoised image, respectively.  Under this metric, high IOU scores close to $1$ measure good agreement/high overlap between the sets while numbers closer to $0$ indicate that the sets are very different. The numerical results are shown in Table~\ref{tab:numerical experiments}(a).   When noise density is less than $0.3$, the IOU scores are above $0.9$, indicating high agreement between the image produced by Algorithm~\ref{Alg:denoise} and the ground truth image.  For noise density $0.4 \sim 0.6$, the IOU are still good for recognizing the ground truth image. When the noise density $\geq 0.7$, although the IOU is below $0.6$, some key features of the image are still visible (see, \textit{e.g.}, Figure \ref{fig:Character}(f)). We also list the result of denoise CNN provided by Matlab 2019, specifically the {\tt{denoiseImage}} function, designed to denoise images using a deep neural network.  The IOU scores are lower for {\tt{denoiseImage}} than for our proposed method. As a final measurement, we also look at the Betti numbers for denoised images, comparing these numbers to the values $(\beta_0, \beta_1)=(6,5)$ for the ground truth image.  These are shown in Table \ref{tab:numerical experiments}(b) and (c). {We observe that Algorithm~\ref{Alg:denoise} produces images with fairly accurate Betti numbers, even for images corrupted by salt and pepper noise at densities up to 0.5 (see middle column of Table~\ref{tab:numerical experiments}(b) and (c)). }

\begin{algorithm}[H]
    \SetAlgoLined
    {\tt{Input}:} $g$, a grayscale image.

    {\tt{Output}:} $\widehat{g}$, denoised grayscale image.
            
    For each $i=0,1,\cdots,255$, calculate $\widehat{g}_i$ by Algorithm \ref{Alg:denoise}.
            
    \tt{Return}: $\widehat{g}:=\sum_{i=0}^{255}\widehat{g}_i$.
            
    \caption{Grayscale image extension of Algorithm \ref{Alg:denoise}.} 
    \label{alg:grayscale ext}
\end{algorithm}

\begin{algorithm}[H]
    \SetAlgoLined 
    {\tt{Input}:} $G=(G^r, G^g, G^b)$, a RGB color image.
    
    {\tt{Output}:} $\widehat{G}$, denoised color image.
    
    For each color channel, calculate $\widehat{G}^r$, $\widehat{G}^g$, and $\widehat{G}^b$ by Algorithm \ref{alg:grayscale ext}.
    
    \tt{Return}: $\widehat{G} = (\widehat{G}^r, \widehat{G}^g, \widehat{G}^b)$.
    \caption{RGB color image extension of Algorithm \ref{Alg:denoise}.}
    \label{alg:color ext}
\end{algorithm}

\begin{figure}
    \centering
    \subfloat[Ground truth.]{
    \includegraphics[width=0.24\linewidth]{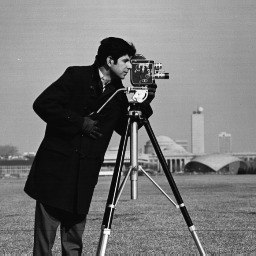}}
    \subfloat[Ground truth.]{
    \includegraphics[width=0.24\linewidth]{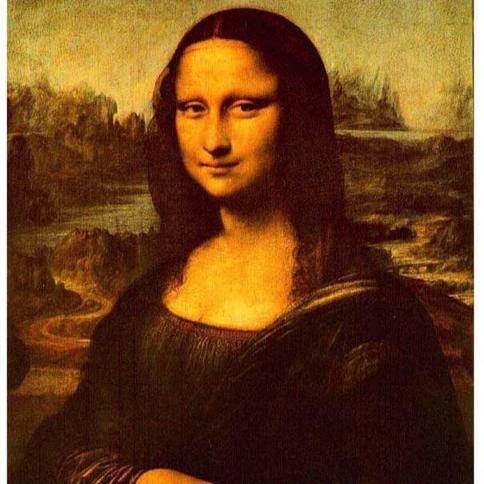}}\\
    \subfloat[Pepper/Salt noise.]{
    \includegraphics[width=0.24\linewidth]{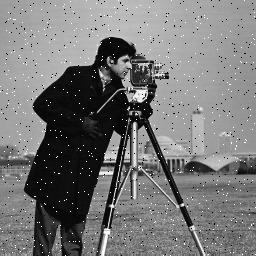}}
    \subfloat[Algorithm \ref{alg:grayscale ext}.]{
    \includegraphics[width=0.24\linewidth]{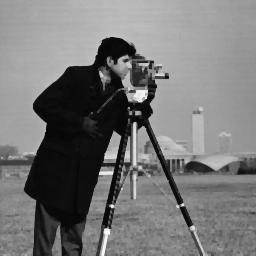}}
    \subfloat[{\tt{denoiseImage}}.]{
    \includegraphics[width=0.24\linewidth]{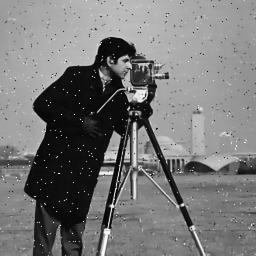}}
    \subfloat[Topaz AI.]{
    \includegraphics[width=0.24\linewidth]{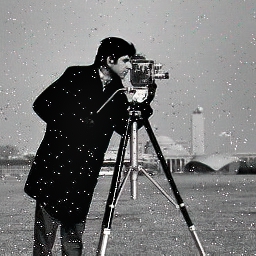}}\\
    \subfloat[Pepper/Salt noise.]{
    \includegraphics[width=0.24\linewidth]{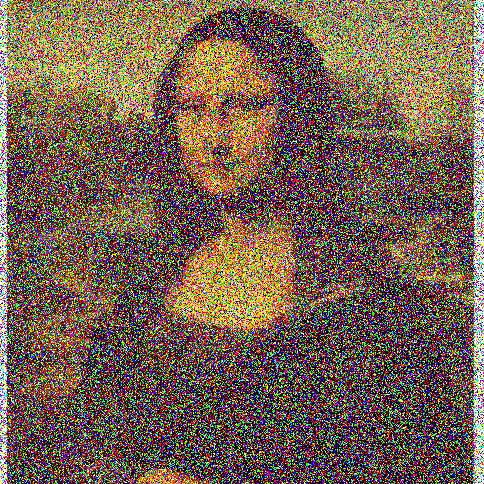}}
    \subfloat[Algorithm \ref{alg:color ext}.]{
    \includegraphics[width=0.24\linewidth]{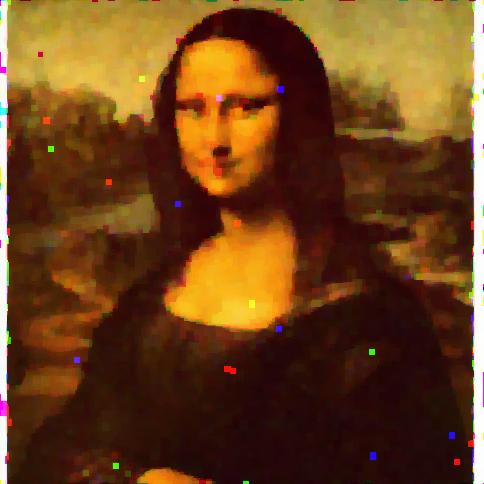}}
    \subfloat[{\tt{denoiseImage}}.]{
    \includegraphics[width=0.24\linewidth]{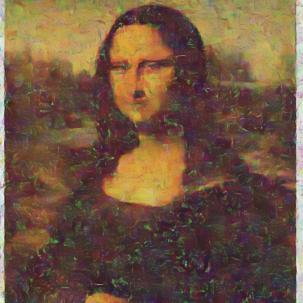}}
    \subfloat[Topaz AI.]{
    \includegraphics[width=0.24\linewidth]{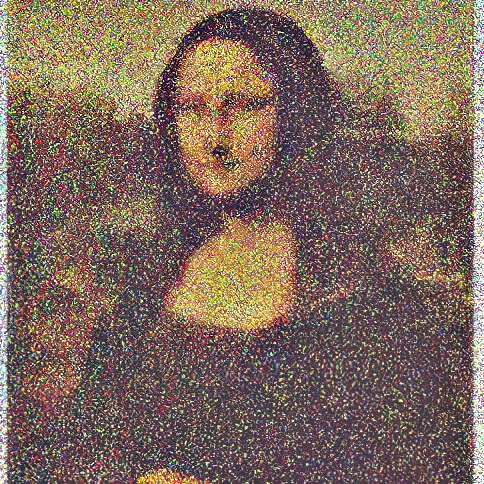}}    
    \\
    \subfloat[{Extended IOU scores.}]{    \begin{tabular}{c|cccc}
    Image & Noise & Algorithm \ref{Alg:denoise}  & {\tt{denoiseImage}} & Topaz AI Denoise \\ \hline\hline
	Figure \ref{fig:gray-scale image experiment}(a) & 0.73 & 0.82 & 0.79 & 0.7412 \\
	Figure \ref{fig:gray-scale image experiment}(b) & (0.55,0.45,0.29) & (0.88,0.86,0.81) & (0.65,0.59,0.52) & (0.51,0.39,0.23)\\ \hline
	\end{tabular}}
    \caption{First row: (a)-(b) ground truth images. Second row: (c) the image (a) with $0.1$ density salt and pepper noise; (d) the denoised image by Algorithm~\ref{alg:grayscale ext} with parameters {\tt{MaxIter}=10} and {\tt SizeTol=4}; (e)  the denoised image by {\tt{denoiseImage}}; (f) the denoised by the Topaz AI Software.  Third row: (g) the image (b) with $0.5$ density salt and pepper noise (for each RGB channel); (h) the denoised image by Algorithm~\ref{alg:color ext} with parameters {\tt{MaxIter}=10} and {\tt SizeTol=7}; (i) the denoised image by {\tt{denoiseImage}}; (j) the denoised image by the Topaz AI Software.  Fourth row: {(k) Average extended IOU scores for 5(a) and by color channel for 5(b). }}
    \label{fig:gray-scale image experiment}
\end{figure}

Combining Algorithm \ref{Alg:denoise} with the thresholding techniques in Proposition \ref{Proposition : Ristriction and opening/closing are commutative} extends this approach to grayscale images. For a grayscale image $g : P \rightarrow \{ 0,1, ..., 255\}$, consider its binary images via global thresholding \eqref{Equation : Grayscale to binary with threshold t}: $g_0, g_1,..., g_{255}$.  We apply Algorithm \ref{Alg:denoise} to each binary image $g_i$ and obtain a denoised binary image $\widehat{g_i}$. The final output grayscale image would be the sum of $\sum_{i=0}^{255}\widehat{g_i}$ {as shown in Algorithm~\ref{alg:grayscale ext}}.  We apply {Algorithm~\ref{alg:grayscale ext}} to a grayscale image as shown in Figure \ref{fig:gray-scale image experiment}(c), and the denoised image is shown in Figure \ref{fig:gray-scale image experiment}(d).  We observe that in Figure \ref{fig:gray-scale image experiment}(d), the salt and pepper noise is removed, and some portions of images (e.g. the mouth of the man, the camera) are blurry.  Note that since each thresholded image, $g_i$, is treated separately by Algorithm~\ref{Alg:denoise}, the resulting images $\widehat{g_i}$ may not form a filtration, i.e.  $\widehat{g_i}^{-1}(0) \nsubseteq \widehat{g_j}^{-1}(0)$ for $i\leq j$.  It would be interesting to extend Algorithm \ref{Alg:denoise} to an approach that would preserve the subset relations on the denoised images, ensuring that they form a filtration.  As a comparison,  the denoised images by {\tt denoiseImage} and Topaz AI are shown in Figure~\ref{fig:gray-scale image experiment}(e) and (f), respectively.  Observe that those images are pixelated.

Using a similar procedure, we may also extend the Algorithm~\ref{Alg:denoise} approach to RGB color images.  Here, we treat each of the three color channels as a grayscale image and follow the same procedure described for grayscale images above as shown in Algorithm~\ref{alg:color ext}.  The results for the 3 channels are then viewed together as an RGB image. Figure~\ref{fig:gray-scale image experiment} (g) shows the constructed noisy images and Figure \ref{fig:gray-scale image experiment} (h) is the image produced by Algorithm \ref{alg:color ext}.  We observe that almost all the salt and pepper noises are removed, and moreover, the denoised image still preserves the original image well. We compare our method which is unsupervised with the commercial denoising software: Topaz AI{~\cite{TopazAILab}} which uses deep learning.  The results are shown in Figure~\ref{fig:gray-scale image experiment}(h), (i) and (j). %We visually observe that images produced by Algorithm \ref{Alg:denoise} are preferable to those by Topaz AI.  
Figure~\ref{fig:gray-scale image experiment}(k) shows the average extended IOU scores, {where the extended score for each grayscale image is the average of the IOU scores over each threshold.  For a color image, the extended IOU score is given for each of the three color channels.}  We observe that Algorithms~\ref{alg:grayscale ext} and \ref{alg:color ext} outperform Topaz AI on these examples.

Our focus in this section has been to demonstrate that the multiparameter filtration contains useful information and that automation may be used to extract it.  Our proposed denoising algorithm works well on binary, grayscale, and color images with salt and pepper noise where the separation in spatial scale between the noise and true features may be used to effectively remove noise.  Recently, others have developed salt and pepper denoising algorithms using deep learning e.g. \cite{8768516,BoFu}, where a training process is required.  Our unsupervised approach does not require training and it would be interesting to investigate whether combining the two approaches could lead to even better results.

\begin{table}
	\caption{Performance of Algorithm \ref{Alg:denoise} for denoising 2D binary images formed from Figure \ref{fig:Character}(a) with added salt and pepper noise. For each prescribed noise density, 1000 images were formed, Algorithm \ref{Alg:denoise} with {\tt{MaxIter}=10} and {\tt SizeTol=5} was applied to each, as was {\tt{denoiseImage}}, a built-in Matlab function for denoising the image by using the deep neural network.  Finally, calculated the IOU scores with respect to the original image (Figure \ref{fig:Character}(a)) was calculated.  IOU scores are listed in the columns below. All scores are recorded by mean $\pm$ standard deviation for the $1000$ trials. (b) Betti numbers by mean $\pm$ standard deviation for all trials.  The Betti pair $(\beta_0, \beta_1)$ of the original image (Figure \ref{fig:Character}(a)) is $(6,5)$.
	\label{tab:numerical experiments}}
\centering
\subfloat[IOU Scores. ]{
\fbox{
		\begin{tabular}{c|ccc}
			 Density   & Noised images  & Algorithm \ref{Alg:denoise} & {\tt{denoiseImage}}  \\ \hline\hline
		    0.1  & 0.8104 $\pm$ 0.0036 & 0.9810 $\pm$ 0.0028 & 0.9311 $\pm$ 0.0042 \\ \hline
		    0.2  & 0.6694 $\pm$ 0.0039 & 0.9551 $\pm$ 0.0106 & 0.9330 $\pm$ 0.0052 \\ \hline
			0.3  & 0.5603 $\pm$ 0.0038 & 0.9139 $\pm$ 0.0202 & 0.8954 $\pm$ 0.0061 \\ \hline
			0.4  & 0.4737 $\pm$ 0.0035 & 0.8825 $\pm$ 0.0313 & 0.8091 $\pm$ 0.0080 \\ \hline
			0.5  & 0.4028 $\pm$ 0.0033 & 0.8345 $\pm$ 0.0145 & 0.6697 $\pm$ 0.0091 \\ \hline
			0.6  & 0.3442 $\pm$ 0.0029 & 0.7323 $\pm$ 0.0171 & 0.5090 $\pm$ 0.0083 \\ \hline
			0.7  & 0.2948 $\pm$ 0.0027 & 0.5802 $\pm$ 0.0180 & 0.3755 $\pm$ 0.0063 \\ \hline
			0.8  & 0.2523 $\pm$ 0.0026 & 0.4194 $\pm$ 0.0147 & 0.2854 $\pm$ 0.0044 \\ \hline
			0.9  & 0.2158 $\pm$ 0.0024 & 0.2961 $\pm$ 0.0108 & 0.2263 $\pm$ 0.0034 \\ \hline
			1.0  & 0.1836 $\pm$ 0.0022 & 0.2162 $\pm$ 0.0100 & 0.1835 $\pm$ 0.0029 \\ \hline
		\end{tabular}	
	}
}\\
\subfloat[$\beta_0$.]{
\fbox{
		\begin{tabular}{c|ccc}
			 Density   & Noised images  & Algorithm \ref{Alg:denoise} & {\tt{denoiseImage}}  \\ \hline\hline
		    0.1  & 247.70 $\pm$ 13.12 & 6.00 $\pm$ 0.06 & 34.14 $\pm$ 6.14 \\ \hline
		    0.2  & 376.19 $\pm$ 13.71 & 5.99 $\pm$ 0.24 & 21.67 $\pm$ 4.21 \\ \hline
			0.3  & 415.75 $\pm$ 14.98 & 5.89 $\pm$ 0.54 & 33.10 $\pm$ 5.40 \\ \hline
			0.4  & 387.48 $\pm$ 15.92 & 5.65 $\pm$ 0.79 & 56.59 $\pm$ 7.22 \\ \hline
			0.5  & 317.93 $\pm$ 17.16 & 5.32 $\pm$ 0.90 & 82.08 $\pm$ 8.45 \\ \hline
			0.6  & 233.26 $\pm$ 16.43 & 5.30 $\pm$ 1.39 & 89.43 $\pm$ 8.94 \\ \hline
			0.7  & 160.11 $\pm$ 14.27 & 7.35 $\pm$ 2.18 & 75.76 $\pm$ 8.99 \\ \hline
			0.8  & 114.04 $\pm$ 11.52 & 17.81 $\pm$ 3.95 & 61.81 $\pm$ 8.38 \\ \hline
			0.9  & 106.59 $\pm$ 11.41 & 47.61 $\pm$ 6.71 & 61.48 $\pm$ 8.57 \\ \hline
			1.0  & 143.97 $\pm$ 12.94 & 70.05 $\pm$ 12.61 & 87.55 $\pm$ 10.07 \\ \hline
		\end{tabular}	
	}
}\\	
\subfloat[$\beta_1$.]{
\fbox{
		\begin{tabular}{c|ccc}
			 Density   & Noised images  & Algorithm \ref{Alg:denoise} & {\tt{denoiseImage}}  \\ \hline\hline
		    0.1  & 1152.40 $\pm$ 29.27 & 4.95 $\pm$ 0.26  & 303.55 $\pm$ 24.25 \\ \hline
		    0.2  & 2050.40 $\pm$ 35.29 & 4.60 $\pm$ 0.72 & 148.82 $\pm$ 17.60 \\ \hline
			0.3  & 2702.80 $\pm$ 37.78 & 4.30 $\pm$ 1.03 & 172.90 $\pm$ 17.26 \\ \hline
			0.4  & 3127.60 $\pm$ 37.49 & 4.97 $\pm$ 1.46 & 323.35 $\pm$ 22.44 \\ \hline
			0.5  & 3343.10 $\pm$ 41.62 & 7.45 $\pm$ 2.37 & 659.75 $\pm$ 33.05 \\ \hline
			0.6  & 3375.70 $\pm$ 45.27 & 16.04 $\pm$ 4.01 & 1195.80 $\pm$ 45.37 \\ \hline
			0.7  & 3257.80 $\pm$ 50.89 & 23.60 $\pm$ 5.43 & 1174.20 $\pm$ 56.11 \\ \hline
			0.8  & 3016.90 $\pm$ 57.08 & 15.95 $\pm$ 3.86 & 2155.30 $\pm$ 56.64 \\ \hline
			0.9  & 2680.30 $\pm$ 59.83 & 3.83 $\pm$ 2.09 & 2229.10 $\pm$ 56.98 \\ \hline
			1.0  & 2284.60 $\pm$ 59.12 & 0.38 $\pm$ 0.66 & 2027.60 $\pm$ 59.26 \\ \hline
		\end{tabular}	
	}
}\\	
\end{table}

\section{Conclusion}
\label{sec:conclusions}

In this work, we establish that, under mild conditions, the morphological operations of erosion, dilation, opening, and closing may be combined to form multiparameter filtrations useful for studying binary images.  These operations may also be combined with thresholding to form yet larger multiparameter filtrations useful for studying grayscale, and, by extension, color images.  The dimension of the filtration grows with the number of operations and structuring elements, forming a potentially high dimensional framework in which to explore image structure and features.  As demonstrated in Sections~\ref{sec:denoising algorithm}, this framework can be used to create automated approaches to image analysis and processing, in our example application leading to methods for removing salt and pepper noise from images.

There is a much broader class of methods for extracting information from multiparameter filtrations than just the approach of calculating persistence along nondecreasing paths that we describe in Definition~\ref{def:nondecreasing} and use in Section~\ref{sec:denoising algorithm}.  Persistent homology may be generalized as a cellular sheaf defined on a partially ordered set $(P,\leq)$, that is, a functor from $P$ to the category of vector spaces~\cite{justinCurryPHDThesis,Curry2015,ghrist2020cellular,robinson2014topological}. Cellular sheaves were originally developed for studying nerve theory in topology~\cite{AllenShepard} and have recently been used for describing the persistence of objects in applied topology. Because the order in Definition \ref{def:multi filtration} is also a partial order on $\bbZ^k$, persistent homology defined on a multifiltration has a natural cellular sheaf structure. The persistence of the structure is much more complicated since the totally ordered property fails on the new order.  However, we do see a variety of approaches for analyzing topological features in this setting, such as sheaf cohomology~\cite{Curry2015,robins2011theory,RobinsonNyquist}, zig-zag homology~\cite{carlsson2010zigzag}, multi-graded Betti numbers~\cite{lesnick2020computing}, and rank invariants~\cite{carlsson2009theory}. The multiparameter filtration we create here offers a constructive class of examples on which to explore these methods.

\bibliographystyle{unsrt}  
\bibliography{references}

\end{document}